\documentclass{article}
\usepackage{natbib,latexsym,epsfig,emulateapj,onecolfloat,graphics}
\def\eps@scaling{.95}
\def\epsscale#1{\gdef\eps@scaling{#1}}

\def\plotone#1{\centering \leavevmode
    \epsfxsize=\eps@scaling\columnwidth \epsfbox{#1}}

\def\plotfiddle#1#2#3#4#5#6#7{\centering \leavevmode
\vbox to#2{\rule{0pt}{#2}}
\includegraphics{#1}}

\def\kms{\ifmmode {\rm\,km\,s^{-1}}\else
    ${\rm\,km\,s^{-1}}$\fi}
\def\ms{\ifmmode {\rm\,m\,s^{-1}}\else
    ${\rm\,m\,s^{-1}}$\fi}
\def\kmsMpc{\ifmmode {\rm\,km\,s^{-1}\,Mpc^{-1}}\else
    ${\rm\,km\,s^{-1}\,Mpc^{-1}}$\fi}
\def\hkmsMpc{\ifmmode {\rm\,h^{-1}\,km\,s^{-1}\,Mpc^{-1}}\else
    ${\rm\,h^{-1}\,km\,s^{-1}\,Mpc^{-1}}$\fi}
\def\lya{{\rm Ly}$\alpha$}

\def\kpc{{\rm\,kpc}}

\def\msun{\ifmmode {\rm\,M_\odot}\else ${\rm\,M_\odot}$\fi}
\def\Msun{\ifmmode {\rm\,M_\odot}\else ${\rm\,M_\odot}$\fi}
\def\lsun{\ifmmode {\rm\,L_\odot}\else ${\rm\,L_\odot}$\fi}
\def\Lsun{\ifmmode {\rm\,L_\odot}\else ${\rm\,L_\odot}$\fi}
\def\rsun{\ifmmode {\rm\,R_\odot}\else ${\rm\,R_\odot}$\fi}
\def\Rsun{\ifmmode {\rm\,R_\odot}\else ${\rm\,R_\odot}$\fi}

\def\cmtw{\ifmmode {\rm\,cm^{-2}}\else ${\rm\,cm^{-2}}$\fi}
\def\cmthr{\ifmmode {\rm\,cm^{-3}}\else ${\rm\,cm^{-3}}$\fi}

\def\ergps{\ifmmode {\rm\,erg\,s^{-1}}\else ${\rm\,erg\,s^{-1}}$\fi}
\def\ergpscmtw{\ifmmode {\rm\,erg\,s^{-1}\,cm^{-2}}}

\def\eg{{\it e.g.}}

\def\deg{\ifmmode {^{\circ}}\else {$^\circ$}\fi}
\def\degr{\ifmmode {^{\circ}}\else {$^\circ$}\fi}
\def\degs{\ifmmode {^{\circ}}\else {$^\circ$}\fi}

\def\etal{{\it et al.~}}

\def\Ho{\ifmmode {\rm\,H_\circ}\else ${\rm\,H_\circ}$\fi}
\def\hnot{\ifmmode {\rm\,H_\circ}\else ${\rm\,H_\circ}$\fi}
\def\h0{\ifmmode {\rm\,H_\circ}\else ${\rm\,H_\circ}$\fi}
\def\hnotunit{\ifmmode {\rm\,km\,s^{-1}\,Mpc^{-1}}\else
    ${\rm\,km\,s^{-1}\,Mpc^{-1}}$\fi}
\def\qnot{\ifmmode {\rm\,q_\circ}\else ${\rm q_\circ}$\fi}
\def\q0{\ifmmode {\rm\,q_\circ}\else ${\rm q_\circ}$\fi}
\def\ie{{\it i.e.}}

\def\vs{{\it versus} }

\def\arcsec{\ifmmode {^{\prime\prime}}\else $^{\prime\prime}$\fi}
\def\asec{\ifmmode {^{\prime\prime}}\else $^{\prime\prime}$\fi}
\def\arcmin{\ifmmode {^{\prime}}\else $^{\prime}$\fi}
\def\amin{\ifmmode {^{\prime}}\else $^{\prime}$\fi}

\def\hone{H{\small I}}
\def\h{{\rm h}}
\def\lesssim{\mathrel{\hbox{\rlap{\hbox{\lower4pt\hbox{$\sim$}}}\hbox{$<$}}}}
\def\gtrsim{\mathrel{\hbox{\rlap{\hbox{\lower4pt\hbox{$\sim$}}}\hbox{$>$}}}}
\let\la=\lesssim                        
\let\ga=\gtrsim

\def\be{\begin{equation}}
\def\ee{\end{equation}}

\begin{document}
\twocolumn
[

\title{The Search for Intergalactic Hydrogen Clouds in Voids}
\author{Curtis V. Manning}
\affil{Astronomy Department, University of California, Berkeley, CA
94720} \authoremail{cmanning@astro.berkeley.edu}

\begin{abstract}


I present the results of a search for intergalactic hydrogen clouds in
voids.  Clouds are detected by their \hone\ \lya\ absorption lines
in the HST spectra of low-redshift AGN.  The parameter with which the
environments of clouds are characterized is the tidal field, for this
places a lower limit on the cloud mass-density which is dynamically
stable against disruption.  The summing of the tidal fields along these
lines of sight is managed by the use of galaxy redshift
catalogs.  The analytical methodology employed here has been designed
to detect gas clouds whose expansion following reionization is
restrained by dark matter perturbations.  The end-products of
the analysis of data are the cloud equivalent width distribution
functions (EWDF) of catalogs formed by sorting clouds according to
various tidal field upper, or lower limits.

Cloud EWDFs are steep in voids ($d \log({d{\cal N}/dz})/d \log({\cal
W}) = {\cal S} \sim -1.5 \pm 0.2$), but flatter in high tidal field
zones (${\cal S} \sim -0.5 \pm 0.1$).  Most probable cloud Doppler
parameters are $\sim 30$ \kms\ in voids and $\sim 60$ \kms\ in
proximity to galaxies.  In voids, the cumulative line density at low
EW (${\cal W} \simeq 15$ m\AA) is essentially equal to that of the
mean EWDF, $\sim 500$ per unit redshift.  The void filling factor is
found to be $0.87 \le f_v \le 0.94$.  The void EWDF is remarkably
uniform over this volume with a possible tendency for more massive
clouds to be in void centers.  The size and nature of the void cloud
population suggested by this study is completely unanticipated by the
results of published 3-D simulations, which predict that most clouds
are in filamentary structures around galaxy concentrations, and that
very few observable absorbers would lie in voids.  Strategies for
modeling this population are briefly discussed.
\end{abstract}

   ]


\section{Introduction}

Suppose we divide the universe into two complementary parts; one
galaxy--dominated, and the other what might be called
space--dominated, or void.  While we know the primary constituents of
galaxy dominated space (hereafter, GDS), little is known of voids.
What lies in voids?  The presence of \lya\ absorbers in low--redshift
spectra raises the possibility that some may be in voids.  What is the
nature of \lya\ absorbing clouds, and how does their nature change as
one looks from void to GDS?  What might be the long-term physical
interaction between voids and the GDS?  What remnants of primordial
perturbations remain in voids to this day?  What is the indicative
total mass density in voids?  How does this compare to predictions
based on numerical simulations?  These are questions which motivated
the present investigation.  Let us take a look at the clouds of the
local universe; let us see what they are doing, with particular
attention to those which are isolated.

\subsection{Clouds Observed}
To begin the study of the effect of environment on the nature of
intergalactic hydrogen clouds one needs the equivalent width (EW) of
their absorption at the \lya\ resonance $(\lambda_{\alpha}=$1215.67
\AA), and their line profiles, characterized by the Doppler parameter
$b$ (\kms), needed to calculate the column density of \hone\ from the
equivalent width.  In addition, we shall require a parameter that will
discriminate between void, and GDS clouds (to be discussed in \S2).

The cloud population is characteristically measured by its line
density $d{\cal N}/dz$, where ${\cal N}$ is the number of clouds with
column density greater than some $N_{HI}$.  The line density may vary
with redshift.  This evolution has been expressed as a power of
$(1+z)$; $d{\cal N}/dz=(d{\cal N}/dz)_0 (1+z)^{\gamma}$
(\citealt{Sargent:80}, hereafter SYBT), where $\gamma$ is sometimes
known as the evolution parameter.  Before the Hubble Space Telescope
(HST) made observation of low-redshift \lya\ clouds possible, our
ideas about the local population of absorbers were based on
extrapolations from the observations of highly redshifted absorption
spectra.  These studies showed a strong evolution of the line density
$d {\cal N}/dz$ with time\footnote{ \citet{Lu:91, Kim:97} show $\gamma
\simeq 2.75 \pm 0.29$ and 2.78 $\pm 0.71$ respectively, though
\citet{Bechtold:94} found a slower evolution $\gamma \simeq 1.89 \pm
0.28$.}.

A series of papers from the HST Quasar Absorption Line Key Project
provided information on clouds at redshifts $z \la 1.5$ using the
Faint Object Spectrograph (FOS)
(\citealt{Bahcall:91,Bahcall:92,Bahcall:93,Bahcall:96,Weymann:98}).
These papers provided palpable proof that the evolution rate of \lya\
clouds at low redshift is substantially lower than at higher redshift.
\citet{Weymann:98} show that for clouds with no associated metal-line
absorbers, the data is consistent with $\gamma=0$ at $z < 1.5$.  For
\hone\ clouds with associated metal lines, $\gamma$ is in the range of
1.24 to 1.55, depending on sample selection criteria.  These absorbers
have been interpreted as arising from galactic halos (SYBT).

\citet{Weymann:98} demonstrated that at low redshifts $\gamma$
declines with EW (see their Fig. 7) indicating that low-EW clouds are
positively evolving (\ie, becoming more numerous) relative to larger
absorbers.  \hone-only clouds with large EW ($\ga 400$ m\AA) were
found to have evolution parameters $\gamma \simeq 0.6$, while the
smaller clouds with ${\cal W}=350$ to 100 m\AA\ have $\gamma \sim 0.5$
to $-0.5$, respectively.  The evolution parameter of very low EW
clouds at low redshift is still unknown, but the large redshift
frequency of low EW absorbers found in \citealt*{Penton:00a}
(hereafter PSSI) suggests that the correlation between $\gamma$ and EW
may continue as the EW approaches 10 m\AA.

The HST has brought the study of \lya\ absorbers
down to a redshift at which cloud environments might be studied.  An
early study of the sightline to 3C 273 (\citealt{Morris:93}) pointed
out the tendency of clouds to avoid clusters of galaxies, and
commented on the existence of one large but very isolated cloud
\footnote{The cloud in question is the highest wavelength cloud (point
with error-bar) in Fig. 6a, in the sightline to 3c273, with an
equivalent width of 297 m\AA}.  Subsequently, other groups
(\citealt{Lanzetta:95}, \citealt*{Tripp:98}, \citealt*{Chen:98})
investigated the proximity of clouds to galaxies, finding an
anticorrelation between cloud equivalent width and the projected
radius to the nearest galaxy.  The same relationship was found to hold
out to $z \approx 0.8$ \citep{Chen:98,Chen:01}.  \citet{Tripp:98}
found that the anticorrelation may extend to projected radii of $\sim
2 $ Mpc from the nearest galaxy, suggesting that galaxies may affect
cloud distributions at very large galactocentric distances.
\citet{Tripp:98} claimed that clouds beyond this radius had a
distribution of equivalent widths similar to those clustered around
galaxies, but that ``void'' clouds appear to be generally smaller.
While galactic impact parameters within $\sim600$ kpc were found to
reliably produce absorption systems, outside this radius, increasingly
frequent 'misses' were recorded, implying a declining covering factor
with increasing radius at a given sensitivity limit.%

The view that these clouds are physically associated with galaxies is
challenged by \citealt{Dave:99}.  They claim that the absorption
systems are associated with uniformly dense filamentary structures
seen in 3-D hydrodynamical simulations, rather than self-gravitating
clouds physically associated with galaxies and groups of galaxies.
Whether associated with galaxies or filaments, there does appear to be
a correlation showing the effects of proximity to galaxies.  The large
physical extent of the radial trend in EWs noted by \citet{Tripp:98},
and the impression of an increasing cloud number density toward
galaxies, suggests that the surrounding void edges are the source of
clouds which are falling into the gravitational potential wells of
galaxies and groups of galaxies.  A very similar distribution of
clouds is detected out to redshifts $z \simeq 0.89$
\citep{Chen:98,Chen:01}, suggesting that such infall is a long-term
feature of the universe.  Other lines of evidence favoring a
substantial accretion rate of centrally condensed clouds are to be
found in the roughly steady, or increasing star formation rate in the
Galaxy disk \citep{RochaPinto:98}, the large dispersion of
metallicities of disk stars as a function of the epoch of their
formation \citep{Edvardsson:93,Carraro:98}, photometric asymmetries
and other signs of accretion of low-metallicity gas in field spirals
\citep*{Zaritsky:95,Zaritsky:97, Haynes:01}, the predominance of
negative barocentric velocities for high velocity clouds (HVC) in the
Local group \citep{Blitz:99,Wakker:99}, and the existence of a
``compact'' subclass of infalling ($\sim 100$ barocentric \kms) HVCs
with cold, compact cores and ionized halos \citep{Braun:00}.

\subsection{Hydro Simulations}
The current view of the nature of the structure and content of the
universe at low redshift is strongly affected by results of numerical
cosmological simulations.  Various 3-dimensional simulations have been
used to model the higher redshift \lya\ cloud population \citep[\eg,][]
{Hernquist:96,Mucket:96,Dave:99,Cen:99}, and some of them have been
extended to low redshift (\citealp*{Riediger:98};
\citealp*{Theuns:98a}; \citealp{Dave:99}).  These simulations suggest
that the clouds with columns $\log{N_{HI}} \la 14$ are produced in the
filaments of roughly homogeneous gas, some of which may still be in
partial participation with the Hubble flow
\citep{Riediger:98,Theuns:98a,Dave:99}, whereas the higher column
density clouds are found in close proximity to galaxy halos.

Very few of the simulations concern themselves with voids \emph{per
se}, so some effort must be taken to extract predictions.
\citet{Riediger:98} found it helpful to separate clouds into two
groups -- shocked $P_s$ and unshocked $P_u$.  Shocked clouds result
from convergent gas flows.  It is thought that the unshocked clouds
lie outside of the filamentary structures, and are expanding with the
Hubble flow so that their column densities would decline steeply with
declining redshift ($N_{HI} \propto (1+z)^5$) at higher redshifts
where $J(z)$ is fairly constant \citep{Riediger:98,Dave:99}, so that
this component produces negligible absorption for $z \le 3.0$.
\citet{Cen:99} argue that from 1/2 to 2/3 of all baryons are in the
filamentary structures constituting a warm to hot medium ($10^5$ K to
$10^7$ K) with a volume filling factor $\sim 0.1$.  This is the
population referred to by \citet{Riediger:98} as ``shocked'', and by
\citet{Dave:99} as producing columns $\log{N_{HI}} \la 14 \, (\cmtw)$
at low redshift.  Void clouds are not expected to produce any
significant \lya\ absorption (Dav\'e, priv. comm. 2001).  However, a
simulation by \citet{Cen:99} finds that about 26\% of baryons are
located in a ``warm'' medium ($T < 10^5$ K) with a filling factor of
0.9 -- this would be the ``unshocked'' population $P_u$ whose
contribution to the line density faded rapidly at redshifts less than
3.  Thus, these simulations suggest that the unshocked region is what
we would call ``void'', and has a current density of ${\bar \rho}_U
\approx 0.26$ of the average, and that filaments, with filling factor
0.1, have a density ${\bar \rho}_S \ga 7$ times the average, so that
the void matter density would be less than 4\% of that in the
filaments.  The issue of the void matter density is addressed directly
by \citet{Einasto:94} through simulations of CDM cosmogony, showing
that matter flows out of low density environments and into high
density areas, resulting in an estimated 15\% of matter being located
in present-day voids.  \citet{El-Ad:97b} estimate a void matter
density of 10\% of the mean based on observation of the galaxy content
of nearby voids.

Given the above, it seems fair to conclude that the predictions of
simulations are for isolated (\ie, void) clouds to be evolving
rapidly, and for the line density in voids, at EWs detectable with
current technonogy, to be quite low.  The overwhelming majority of
absorption systems are expected to be produced by gas associated with
the filamentary structures having warm to hot temperatures and a
filament filling factor $f_f \sim 0.1$ \citep{Dave:99}.

\subsection{Structure Formation}
The conclusions from more theoretically-based studies of structure
formation are somewhat different.  In semi-analytic simulations of
very high redshifts (\citealt{Abel:98}; \citealt*{Bromm:99};
\citealt*{ Nakamura:01}), molecular hydrogen, formed shortly after
recombination, provides cooling for low-mass perturbations which
results in the formation of Population III stars at $z \simeq 20$ in
halos with halo velocities as low at $\sim 1.5$ \kms\ suggesting that
even very small halos with masses in the range $10^5$ to $10^9
~\Msun$, would have formed stars.  However, this field currently
suffers from a ``missing halos'' problem; there should be many more
dwarf galaxies within the Galaxy's halo under the favored cold dark
matter cosmogony (CDM) than are in fact observed.  Estimates of the
magnitude of the discrepancy between predicted and observed low-mass
galaxies range from a few \citep{Kauffmann:93} to a factor of $\sim
50$ (\citealt{Klypin:99,Moore:99a}).  It appears that the discrepancy
manifests itself increasingly below halo velocities of $\sim 40$ \kms.
The deficit in numbers between predicted low-mass mini-galaxies, and
that suggested by observations of Local Group galaxies implies that
the overwhelming majority of small halos ($v_c \la 25 ~\kms$), for
whatever reason, did not in fact generally form stars.  Therefore,
these halos could very well be present in the universe today as
clouds.  But while as mini-galaxies they could survive the tidal
fields around large galaxies, as \emph{clouds} the ram pressure of
dissipated gas around galaxies and groups of galaxies would cause the
stripping of halos of their baryons, rendering them invisible.
Therefore, the cloud density may be less than the numbers of
mini-galaxies predicted to have survived to the present with identical
halo velocities due to their disruption in dense environments.  Their
remnants may remain as isolated (\lya\ clouds), or they may be
currently falling into galaxy halos and identified as HVCs as
suggested by \citet{Klypin:99} and \citealt*{Grebel:00}.  Yet it is
claimed \citep{Klypin:99} that their numbers are not inconsistent with
the full deficit of halos.

\subsection{This Paper}
The general success of bottom-up galaxy formation scenarios in
predicting large-scale structure provides considerable
justification for accepting the notion that halos of sub-galactic size
were already in place at the epoch of reionization with their baryons
somewhat relaxed into their halos.  Yet the paucity of mini-galaxies
today suggests that cooling could not generally have proceeded to the
point of star formation, and that somewhat larger clouds, formed later
but with the greater cooling rates afforded by their stronger
self-gravity, formed the first stars.

It is reasonable to suppose that following
reionization, baryons were heated and flowed outward
\citep{Barkana:99}, but that they may still be in the proximity of
those halos.  The dark halos of clouds provide a means to restrain the
dispersal of baryons after the epoch of reionization.  Though the
ability to retain the outflow is dependent on the halo model, this
could help to explain the apparent endurance of low equivalent width
clouds, as suggested by the low observed values of their evolution
parameter \citep{Weymann:98}.

High-sensitivity observations of low-redshift \lya\ clouds may provide
a means to corroborate these assertions.  PSSI used HST/GHRS/G160M to
study the absorption spectra of bright, low redshift Seyfert and BL
Lac galaxies at high spectral resolution ($\sim 19 \,{\rm \kms}$, or
80 m\AA).  Their observations were supplemented by archival data from
the same instrument.  Some of these spectra reached a ``sensitivity''
(see \S4.2) as low as about 12 m\AA.  One of the important outcomes of
their analysis (\citealt*{Penton:00b}, hereafter PSSII) was the
determination of the mean \hone\ equivalent width distribution
function $f({\cal W})$.  Results showed that the redshift frequency of
\lya\ clouds with ${\cal W} \ga 12$ m\AA\ (\hone\ columns $N_{HI} \ga
10^{12.3} \, \cmtw$ for $b\simeq 25 ~\kms$) approach $d {\cal N}/dz
\simeq 450$ per unit redshift.  In their modeling of the clouds, PSSII
suggested that 20-45\% of all baryons in the low redshift universe are
associated with these clouds.

In contrast to expectations based on the simulations mentioned above,
both PSSI and \citet{Tripp:98} show that significant numbers of clouds
(up to $\sim 50\%$) appear to be very isolated.  One wonders, what
exactly is going on in galaxy voids?  This paper is an attempt to
answer that question.  The paper begins with an analysis of the
observational basis for various measurable cloud quantities (\S2),
seeks the proper way to specify the environment of a cloud, and then
shows how galaxy redshift catalogs can be used to measure it (\S3).
The source and nature of the data used in this study are presented in
\S4.  Details on the method by which mass is attributed to galaxies,
and how tidal fields are summed, are presented in \S5.  Section 6
presents the distribution functions of clouds specified by tidal field
upper and lower limits.  In addition, a variational approach is used
to investigate the effects of various assumptions on the output
distribution functions.  There is a discussion (\S7) and conclusions
(\S8).

Though the effect of employing different cosmologies on the results of
this analysis is minimal, a cosmology must be specified, as future
modeling will require a cosmological context. A flat FRW cosmology
with $\Lambda=0.7$, and $h=0.75$ is used, unless otherwise noted.  The
cosmic baryon density is assumed to be $\Omega_b/\Omega_m=0.11$.

\section{About Clouds}

Before studying these clouds in particular, it may be helpful to
review their basic observational relationships.  The cumulative line
density $d{\cal N}/dz$ of evenly distributed clouds of comoving number
density $n$ and which produce a column density of $\ge N_{HI}$ (or,
alternatively, an EW $\ge {\cal W}$) at projected distance $r_{p}$
from the center of a spherical cloud is, \be \frac{d{\cal N}}{dz}={n
\pi r_{p}^2 (1+z)^3}\, \frac{dl}{dz}, \ee where $dl$ is the element of
proper path length, such that the comoving distance coordinate element
$dr=(1+z) \, dl$, and the column density is formally defined by the
integral of the \hone\ density over a column of length ${L}$ through
the cloud at a projected radius $r_p$, \be N_{HI} = \int_{L} n_{HI}(r
\ge r_p) \, d{L}.  \ee We also have, by way of a definition of
$\gamma$, \be \frac{d{\cal N}}{dz}=\left(\frac{dN}{dz}\right)_0
(1+z)^{\gamma}.  \ee Solved for $n$, Eqs. 1 and 3 yield, \be
n=\left(\frac{d{\cal N}}{dz}\right)_0 \frac{(1+z)^{\gamma-3}}{\pi
r_{p}^2} \frac{dz}{dl}. \ee For a flat universe with non-zero
cosmological constant,
\begin{equation}
\frac{dl}{dz} = \frac{R_0}{(1+z) \sqrt{\Omega_m(1+z)^3 + \Omega_{\Lambda})}},
\end{equation}
\citep{Scott:00} where $R_0=c/H_0$.  Assume that the radius at which a given
column density occurs in an individual cloud is 
\be 
r_{p} \propto(1+z)^{\epsilon},
\ee
then taking the derivative of Eq. 4 with respect to
$z$ we find, 
\be
 \frac{1}{n} \frac{dn}{dz} = \frac{\gamma -2 \epsilon
-3}{(1+z)} - \frac{d^2 l/dz^2}{dl/dz}.  
\ee 
The last term is,
\begin{displaymath}
\frac{d^2l/dz^2}{d l/d z} = - \frac{1}{1+z} \left[1 + \frac{3/2}{(1 +
\Omega_{\Lambda}/(\Omega_0(1+z)^3))} \right].
\end{displaymath}
Substitute the above into Eq. 6 to find,
\begin{displaymath}
\frac{1}{n} \frac{dn}{dz} = \frac{1}{1+z}\left[\gamma+ \frac{3/2}{1 +
\Omega_{\Lambda}/(\Omega_0(1+z)^3)} -2 \epsilon - 2\right].
\end{displaymath}
Assume that the comoving number density of some population of clouds is
constant so that $1/n(dn/dz)=0$.  Then,
\begin{equation}
\gamma =  2+ 2\epsilon - \frac{3 (1+z)^3
\Omega_m}{2 \sqrt{\Omega_m (1+z)^3 + \Omega_{\Lambda}}}.
\end{equation}
Let us consider the specific case in which there are associated metal
lines.  It is probable that the radius at which metal lines are
detected is fairly constant if they require a hot halo to maintain
them in ionized form.  In that case, the cross section for absorption
is constant, and $\epsilon=0$.  With $dn/dz = 0$ at $z=0$, we find the
evolution parameter for metal-line clouds is
\begin{displaymath}
\gamma_{metal}=1.55,
\end{displaymath}
for the adopted cosmological parameters (\S1), consistent with the
evolution parameter for metal-line \hone\ clouds observed
by\citet{Weymann:98}.  However, when we are dealing with primordial
\lya\ clouds that are DM-held, the radius at which a given column
density is observed may change.  We decompose $\epsilon$ into the
effects of a changing metagalactic flux and the cloud density profile.
Assume that
\begin{eqnarray}
J(z)&=&J_0(1+z)^{\xi} \\
\rho(r)&=&\rho_0 r^{-\delta}.
\end{eqnarray}
Equation 10 entails a temporally static matter distribution for each
cloud.  This may be unrealistic; small clouds may still be expanding,
and larger clouds may be recollapsing.  However, analytical modeling
at this level of detail is clearly inferior to hydrodynamic
simulations, so we forgo this in favor of the simplification that
clouds are static.  Since optically thin gas has a neutral fraction
inversely proportional to $J_0(z)$, and directly proportional to
$n_H^2$, then $n_{HI} \propto r^{-2 \delta}/(1+z)^{\xi}$.  The
column density at a cloud projected radius $r_p$ is then $N_{HI}
\propto r_p^{-(2 \delta -1)}/(1+z)^{\xi}$.  The impact parameter
$r_p$ at which the cloud has a neutral column $N_{HI}$ varies as, \be
r_p \propto (1+z)^{-\xi/(2 \delta -1)}. \ee Apparently we may
substitute $-\xi/(2\delta-1)$ for $\epsilon$.  If $\xi
\sim 3$ or so, as suggested by the work of \citet{Shull:99}, and
$\delta \sim 1.75$, a somewhat flattened baryon distribution, then
$\epsilon=-1.2$, and at $z=0$, Eq. 8 yields,
\begin{displaymath}
\gamma_{cl} \simeq - 0.85,
\end{displaymath}

If we can extend the $\gamma - EW$ correlation presented in
\citet{Weymann:98} down to, say, 12 m\AA, then the above is in
agreement.

But what are the predictions of simulations?  As \citet{Riediger:98}
point out, if the ionizing flux is approximately constant, then the
column density for clouds expanding with the Hubble flow is $N_{HI}
\propto n_H^2 L \propto (1+z)^5$.  Therefore, if $J_{0} \propto
(1+z)^\xi$, where $\xi \simeq 3.0$, then $N_{HI} \propto (1+z)^2$.  

To summarize, the analytical approximations of \citet{Dave:99} and
\citet{Riediger:98} are based on the presumption of homogeneous clouds
of negligible mass which expand freely but have sharp edges, whereas
the view presented in this paper contends that the clouds are
self-gravitating, and have the gas density gradients characteristic of
self-gravitating clouds.  As suggested by \citet{Dave:99}, their cloud
model can explain the observed low $\gamma$ for low EW clouds by
citing the effects of a decline in flux of the ionizing background
with time, but to accomplish this $\xi$ (Eq. 9) must be of order 4.5,
rather more than suggested by the work of \citet{Shull:99}.  Our
preference is that cloud density gradients in conjunction with a
decline in $J_0(z)$ are responsible for the low observed cloud
evolution parameter \citep{Weymann:98}.

\section{The Environments of Clouds}

Self-gravitating clouds are delicate structures and can be easily
disrupted.  We are searching for one characteristic which by its
absence allows clouds to abide, while in its presence the cloud is
destroyed.  Intuitively, \emph{isolation} corresponds to the lack of
that factor which destroys clouds.  Hence, our parameter deals with
the proximity to galaxies and groups of galaxies.  Heretofore, the
degree of isolation of clouds has been generally characterized by the
distance (in projected radius and velocity) to the nearest galaxy.
However, one megaparsec from a small galaxy is not the same as one
megaparsec from a giant.  Optimally, one should weight the influences
of galaxies by their mass and distance.  Besides the effects of
ionizing radiation, which is here assumed to be uniform over the small
range of redshifts of the data to be studied, the principal
constraints on the physical integrity of clouds is from shear forces
or ram pressure, and the strength of the tidal field.  In addition,
the heightened probability that a small halo would be accreted to a
larger one is a direct function of ambient density; it is expected to
produce an effect similar to that of a tidal field -- a flattening of
the cloud EW distributions in high density areas.  For isolated
clouds, only tidal fields are relevant, though the flux of ionizing
photons may cause heating which will help to disperse the baryonic
content of small clouds.  Because a uniform distribution of mass will
have no tidal effect on a cloud, galaxies are the main sources of
tidal fields on large scales.

\subsection{Tides produced by galaxies}
If one takes the second spatial derivative of the summed gravitational
potentials of galaxies at a particular point, one gets a quantity
indistinguishable from the tidal field.  One can imagine, then, that
minima in the magnitude of tidal fields would occur in galaxy voids.
The tidal field $T$ is in fact a function expressible in units of
inverse time squared, so that if it is multiplied by the Hubble time
squared it becomes a dimensionless quantity ${\cal T}= T/H_0^2$.  The
dimensionless form ${\cal T}$ is hereafter referred to as the ``tidal
field'' parameter.  Tides may be of vector or scalar form.  In this
paper, a scalar tide is calculated, though we must first consider its
vector form.  For an individual galaxy, the only non-zero component of
${\cal T}$ is radial;
\begin{equation}
{\cal T}_R= - \frac{1}{H_0^2}\frac{d}{d \,R}\left(\frac{GM(R)}{R^2} \right).
\end{equation}
In summing the effects of many galaxies at a given point along the
line of sight (LOS), we must decompose this into Cartesian components.  For instance,
the $x$-component, transverse to the LOS, is
\begin{equation}
{\cal T}_x = \frac{2 G M(R)}{H_0^2 R^4} |\Delta x|,
\end{equation}
where $R=\sqrt{\Delta x^2 + \Delta y^2 + \Delta z^2}$.  The components
transverse to the observer's LOS are calculated using the angular
diameter distance pertinent to the cosmological model employed in this
paper.  The proper displacement $\Delta z$ is found using Eq. 5 (\S2).
That is, the radial (LOS) position is determined by the redshift
alone; no allowance is made for peculiar velocities.  This causes some
positional error, which will be discussed in \S5.5.  In this manner,
the contributions of each galaxy can be added, then combined in
quadrature to arrive at an averaged scalar tidal field ${\cal T}$
acting on a cloud,
\begin{equation}
{\cal T}= \frac{\sqrt{{\cal T}_x^2+{\cal T}_y^2+{\cal T}_z^2}}{3}.
\end{equation}

Let us see how the tidal field works to disrupt clouds.  The condition
for stability to tidal disruption is as follows,
\begin{equation}
H_0^2{\cal T} r \leq \frac{G m(r)}{ r^2},
\end{equation}
where $m(r)$ is cloud mass summed to radius $r$.  When a cloud is
close to a single galaxy, then the radial form will be more pertinent.
By application of Eq. 12 to single galaxy, the tidal field is,
\begin{equation}
{\cal T}_R = \frac{GM(R)}{H_0^2 R^3}\left(2-\frac{d~\log{M}}{d~\log{R}}\right).
\end{equation}
For an isothermal halo, ${d~\log{M}}/{d~\log{R}}$ is unity when $R \le
R_t$, and zero beyond that.  The standard halo model is that of
\citealt*{Navarro:96,Navarro:97} (hereafter NFW), which we discuss
in more detail below (\S 5.1.2).  The above derivative for the NFW
halo does not approximate a constant function of $r$ unless baryons
``conspire'' to produce a flat rotation curve inside $r_{max}$.
However, beyond $r_{max}$ the density profile tends toward an index of
$-3$, and a simple solution for ${\cal T}_R$ no longer maintains.

The simplest halo is isothermal, possibly truncated at some
characteristic radius $R_t$.  The mass of such a halo increases
linearly with radius, so that one may describe this radial mass distribution
by a constant, ${\cal K}\equiv M(R)/R$.  The circular velocity of such
a system is $v_c=\sqrt{G {\cal K}}$, and the total mass is
$M_{tot}={\cal K} R_t$.  For a truncated isothermal halo, Eq. 13
yields,
\begin{equation}
{\cal T}_R = \left\{ \begin{array}{ll}
        \frac{G{\cal K}}{H_0^2 R^2}  .............................. (R \leq R_t) \\
        \frac{2 G M_{tot}}{H_0^2 R^3}........................... (R \geq R_t).
       \end{array}
    \right.
\end{equation}
Note that the effect of halo mass truncation on the tidal field is to
produce a discontinuity at $R_t$.  While a sharp
truncation is not realistic, it will not produce a significant effect
on the distributions of clouds as we shall in all cases assume a total
mass.

In the context of a specific cloud, Eq. 15 can be written,
\begin{equation}
\frac{H_0^2 {\cal T}}{G} \le \frac{m}{{r}^3} = \frac{4 \pi}{3} {\bar \rho}(r),
\end{equation}
where ${\bar \rho}(r)$ is the average over-density of the cloud within
radius $r$.  Thus, the requirement for stability can be stated,
\begin{equation}
{\bar \rho} \ge \frac{3 H_0^2{\cal T}}{4 \pi G} = 2 {\cal T} \rho_{crit},
\end{equation}
where $\rho_{crit}$ is the critical density, $\rho_{crit}=3H_0^2/8 \pi
G$.  There is one important caveat to this: the proper number density
of galaxies declines with decreasing redshift due to the expansion of
the universe.  Therefore, since the average distance of galaxies from
an average void cloud is increasing with time, it is clear that in the
past the tidal field on a void cloud varies roughly as ${\cal T}
\propto (1+z)^3$.  However, clouds which are in some proximity to a
galaxy may have a weaker ${\cal T}$ with lookback time if the peculiar
velocity toward the galaxy is greater than the $H R$.  When we think
of the tidal field affecting void clouds we must antedate to some
extent the current ${\cal T}$ to get a feel for what it has been
subjected to in the past.  For instance, at a lookback time of 7.2 Gyr
($z=1$), the tidal field affecting a void cloud will be $\sim 8$
times the present value.

Self-gravitating clouds have negative radial density gradients.  The
${\bar \rho} \ge 2 {\cal T} \rho_{crit}$ criterion applies
differentially to the clouds within which $M(r)/V(r) > 2 {\cal T} \,
\rho_{crit}$.  Though tidal fields affect all clouds, only the
smallest are destroyed.  Thus the effects of tides will be manifested
over the distribution of cloud sizes, and hence over the whole range
of column densities and EWs.  The physical effect will be a tidal
truncation, reducing the average column density a cloud would produce
for a given halo velocity.  It is expected to affect low-mass clouds
more dramatically than more massive clouds.

A cloud near a galaxy will experience a radial tidal field, and it is
unlikely to have experienced a higher tide in the past, given its
probable infall into the galaxy's (or group) potential well.  An example
may clarify the utility of ${\cal T}_R$.  A homogeneous cloud with an
average over-density ${\bar \rho}=2 \rho_{crit}$ would be marginally
destabilized by a tidal field of ${\cal T}_R =1.0$.  A galaxy with a
flat rotation curve of $v_c=220$ km/s would destabilize this cloud at
a distance of 2.08 $h_{75}^{-1}$ Mpc if the galaxy halo truncates at
0.5 $h_{75}^{-1}$ Mpc, or at 2.94 $h_{75}^{-1}$ Mpc if it has an
untruncated isothermal halo.  Thus we would not expect to find
discrete clouds of low EW in close proximity ($\la 2$ Mpc) to
galaxies.  A galaxy with an NFW halo of the same velocity would have a
tidal field ${\cal T}=1.0$ at a distance slightly less than 2/3 times
that of the truncated isothermal halo.


We now turn our attention to the problem of sorting clouds by
their environments.  If the tidal field is calculated along the
lines of sight to AGN, then if one specifies a limiting tide ${\cal
T}$, absorption systems can be separated into two catalogs,
one with ${\cal T}$ providing the upper limit, denoted ${\cal
C}({\cal T}^U)$, and the other with ${\cal T}$ as a lower bound,
${\cal C}({\cal T}^L)$.  The actual calculated value of the tidal
field will depend on which galaxy halo model is used.  The NFW halo is
the outcome of analysis of N-body simulations, and has a ``cuspy''
core approaching a density $\rho \propto r^{-1}$.  However, more
recent, higher-resolution simulations find the core of the halo to be
even more cuspy than the NFW model ($\rho \propto r^{-1.5}$)
\citep{Moore:99b}, suggesting a conflict with observed galaxies
\citep{Firmani:00,Burkert:00,Wu:00}.  Furthermore, the NFW halo may be
inconsistent with what is suggested by the steadiness of the radial
dependency of the velocity dispersions of dwarf galaxies about their
parent giant field galaxies to radii $r_p \approx 500 \, h_{75}^{-1}$
kpc \citep{Zaritsky:94,Zaritsky:97a}, far beyond the radius at which
the halo has a maximum circular velocity ($r_{max}$ is of order 15--40
\kpc).  Because of these perceived problems with the NFW model, and
the simplicity of the truncated isothermal halo, the latter is used as
a standard model in this paper, relegating the NFW to the position of
first alternative model.

It is possible to coarsely see the effect of tides in the original
PSSI data by binning absorption systems according to the
projected radii (with $\Delta v < 500$ \kms) to the nearest galaxy.
The results show that very low column density systems are not seen
within $\sim$ 1 Mpc of a galaxy (Fig. 1, top panel), but the farther
one goes from the nearest galaxy (middle and bottom panels), the lower
the Doppler parameters and column densities which are recorded.  If
void clouds were freely expanding, one would not see lower b-values
with decreasing EWs, as the bulk velocity of expansion is added to the
thermal broadening.

\begin{figure}[h!] 
\centering
\epsscale{0.85} 
\plotone{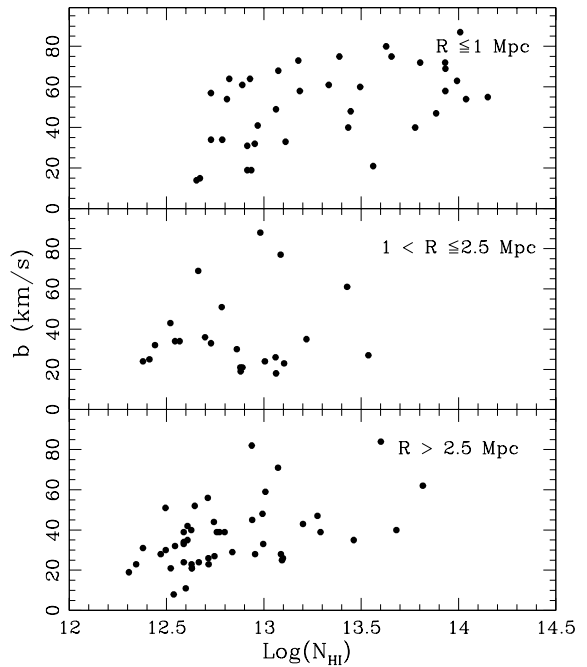}
\caption{Column density as a function of b-value for data from PSSI
for 3 bins of projected radius to the nearest galaxy.  Radius was
derived using a ``retarded'' Hubble flow model (clouds are placed at
the galaxy redshift when they are within $\pm 300$ \kms\ of the
galaxy).  Cloud b-values appear to have lower limits which correlate
inversely with galactocentric radius.  The variation of the lower
limits of column densities with radius are consistent with the effects
of tidal fields.}
\end{figure}

These provisional findings are consistent with the notion that tidal
fields disrupt smaller clouds, and also supports the concept that the
expansion of baryonic clouds are at least restrained by the gravity of
their dark halos.  It is reasonable to suppose that outside of massive
halos, tidal fields provide the main constraint on cloud stability.
We therefore formally adopt the tidal field ${\cal T}$ as the
discriminator for cloud environments; a low ${\cal T}$ then indicates
an environment in which cloud density is essentially unconstrained,
and a high ${\cal T}$ would mean that clouds with density $\rho_{cl}
\la 2 {\cal T} \rho_{crit}$ would be disrupted.

\section{The Data}

The basic data (PSSI) required for calculating the distribution
$f({\cal W})$ as a function of tidal field are of three types; cloud
absorption-line data (wavelength $\lambda$, rest equivalent width
${\cal W}$, Doppler parameter $b$), sensitivity functions, and galaxy
redshift catalogs.  We begin with the HST/GHRS/G160M data of PSSI
along 15 sightlines toward luminous, low-redshift Seyfert and BL Lac
galaxies.  Table one presents 15 target RA and Dec as well as the
spectral range covered by the data for each line of sight.

\begin{table}[h!] 
\begin{center}
\caption{Spectral targets and wavelength range}
\medskip
\scriptsize
\begin{tabular}{lcccc}
\tableline
\tableline

{ Target } & RA & Dec (J2000) & $\lambda_{min}$ (\AA) & $\lambda_{max}$ (\AA) \\

\hline

3C 273 & 12:29:06.7 &  +02:03:08.6 & 1214.11 & 1301.46 \\
Arkelian 120 & 05:16:11.4 & -00:08:59.0 & 1222.51 & 1258.75 \\
ESO 141-G55 & 19:21:14.3 & -58:40:13.0 &  1222.51 & 1265.92 \\
Fairall 9 & 01:23:45.8 &  -58:48:20.0 & 1219.79 & 1276.38 \\
H1821+643 & 18:21:57.3 & +64:20:36.4  & 1231.66 & 1267.75 \\
IZW 1 & 00:53:34.9 & +12:41:36.0 & 1221.40 & 1257.54 \\
Mrk 279 & 13:53:03.4 & +69:18:29.6  & 1222.53 & 1258.74 \\
Mrk 290 & 15:35:52.4  & +57:54:09.2 & 1231.76 & 1268.99 \\
Mrk 335 & 00:06:19.5 & +20:12:10.5 & 1221.41 & 1257.55 \\
Mrk 421 & 11:04:27.3 & +38:12:31.8 & 1221.43 & 1257.57 \\
Mrk 501 & 16:53:52.2 & +39:45:36.6 & 1221.42 & 1257.56 \\
Mrk 509 & 20:44:09.7 & -10:43:25.0 & 1219.47 & 1267.84 \\
Mrk 817 & 14:36:22.1 & +58:47:39.4 & 1222.57 & 1258.77 \\
PKS 2155-304 & 21:58:52.1 & -30:13:32.1 & 1222.58 & 1293.67 \\
Q1230+0155 & 12:30:49.9  & +01:15:23.0 & 1216.96 & 1254.23 \\
\tableline
\end{tabular}
\end{center}
\medskip
\normalsize
\vspace{-1cm}
\end{table}

\subsection{The Absorption Systems}

Source spectra probe sightlines to low-redshift active galaxies
brighter than $V \leq 14.5$ mag (Table 1).  Only two of the sightlines
include data beyond a redshift $z=0.045$.  The well-studied sightline
to 3C 273 includes clouds with the largest maximum redshift of
$z\simeq0.07$.

The data include cloud wavelength and error, equivalent width, and its
error, the resolution-corrected Doppler parameter, and its error.
Some of the spectra are pre--, and some post--deployment of COSTAR
(Corrective Optics Space Telescope Axial Replacement), but there is no
attempt to separate them except insofar as their sensitivity functions
are calculated somewhat differently.  No clear trend in Doppler
parameter is noticed between pre- and post-COSTAR data.

The clouds are considered to be randomly selected from a parent
population -- one which is effected by ambient tidal fields -- within
the constraints of the sensitivity of the instrument along the line of
sight (LOS).  The reader is referred to PSSI for a detailed
description of the reduction of the HST data.  Their procedure
involves fitting the continuum to a polynomial by increasing the order
of the polynomial until no further reductions in $\chi^2$ are
possible.  All negative spectral fluctuations from the continuum
greater than 1-$\sigma$ are treated as the original absorption line
list, and the lines are fitted with gaussian components accepting only
those with values $12 < b_{obs} < 100$ \kms.  The significance levels
of clouds are calculated by integrating the signal to noise ratio per
resolution element over the cloud (see below).

\subsection{Sensitivity functions}

The sensitivity functions indicate what cloud EWs can be observed
given the significance level imposed for the cloud catalog.

Let us begin with a review of the relationships between
signal-to-noise ($S/N$), significance level ($SL$), EW, and
sensitivity (${\cal S}$).  The signal $S$ of an absorption system is
the integral of the counts-deficit relative to the continuum over a
resolution element, \be S = \int_{RE}(f_{cont}(\lambda)-f_{\lambda})
\,S_G \,(t/4) \, d\, \lambda \equiv \Delta n, \ee where $S_G$ is the
GHRS sensitivity at $\lambda$ in units \\ ${\rm counts
\,s^{-1}\,diode^{-1}\,ergs^{-1}\,s^{-1}\,cm^{-2}\,\AA^{-1}}$ (GHRS
handbook), and the absorption system is represented by the difference
of the continuum $f_{cont}$ and the observed flux $f_{\lambda}$.  The
time $t$ is divided by 4 because the exposures are ``quarter stepped''
using FP-SPLIT to help remove the effects of photocathode granularity
and diode to diode detector variations.  The noise $N$, based on the
total counts in the continuum integrated over one RE, is \be
N=\sqrt{\int_{RE} f_{cont} \,S_G \,(t/4) \, d\, \lambda} = \sqrt{n},
\ee so within a given RE, \be S/N=\frac{\Delta n}{\sqrt{n}}. \ee Other
sources of noise are said to be small so that the S/N is within
$\sim5\%$ of the ``root n'' photon statistics (Final Report, GHRS
Science Verification Program, 1992).  It will be assumed that ``root
n'' statistics are accurate though it may be a small under-estimate of
the true sensitivity in m\AA.  The resolution element $RE$ is
estimated at $1.2$ diodes for pre-COSTAR and $1.1$ diodes for
post-COSTAR installation (PSSI).  The observed equivalent width of a
cloud within a resolution element is, \be EW=\frac{\Delta n}{n/RE} =
\frac{\Delta n}{n} RE,\ee where $RE\simeq 80$ m\AA (PSSI).

A cloud detected at a 4.0-$\sigma$ SL in one RE has an EW ${\cal
W}=\Delta n \, RE/n$ such that $\Delta n = 4.0 \sqrt{n}$.  The
sensitivity per RE, ${\cal S}$, is the EW of a 4-$\sigma$ cloud.
Thus,
\begin{equation}
{\cal S}=\frac{4 \sqrt{n}}{n}RE = \frac{4 RE}{\sqrt{n}}
\end{equation}
These sensitivity functions are plotted as solid lines in Fig. 2.

The total cloud EW is the sum of the EWs of each resolution element.
The total significance level of the cloud is the imposed significance
level times the summed equivalent width per RE divided by the
sensitivity per RE.  Using the equations above, terms cancel, and the
SL of the cloud is, \be SL=\sum_i \frac{\Delta n(i)}{\sqrt{n(i)}}.\ee
Thus the SL of the cloud is the sum of the S/N in each RE of the
cloud.  In practice this is done by integrating the $S/N$ over the
best fit Gaussian.  Thus, if a cloud of a given EW is found in one RE,
and if its $EW \ge {\cal S}$, then a cloud with a similar EW, but
spread over more RE's, will have the same significance level, for the
counts deficit per RE $\Delta n \propto b^{-1}$, while the number of
RE's spanned by the cloud is $\propto b$; in combining them the
$b-$factors cancel.  Thus, if a cloud appears above the sensitivity
function, then it is a 4-sigma cloud, regardless of its $b$-value.

\begin{figure}[h!] 
\setcounter{figure}{1}
\centering
\epsscale{1.}
\plotone{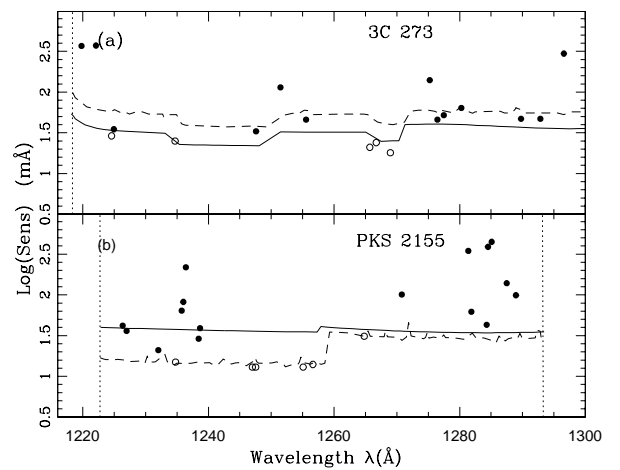}
\vspace{-2.0cm}
\caption{The sensitivity functions (panels a-o) as derived by the
author from the flux (solid line) and extracted from
PSSI on the basis of fluctuations around the continuum.
The dashed lines are the sensitivity functions from
PSSI.  Clouds with significance greater than 4-$\sigma$
(as assigned in PSSI) are represented by solid dots;
open dots are for clouds with SL in the range, 3-$\sigma \leq SL \leq
$4-$\sigma$}
\end{figure}

\begin{figure}[h!] 
\setcounter{figure}{1}
\centering
\epsscale{1.15}
\hspace{-1.5cm}
\plotone{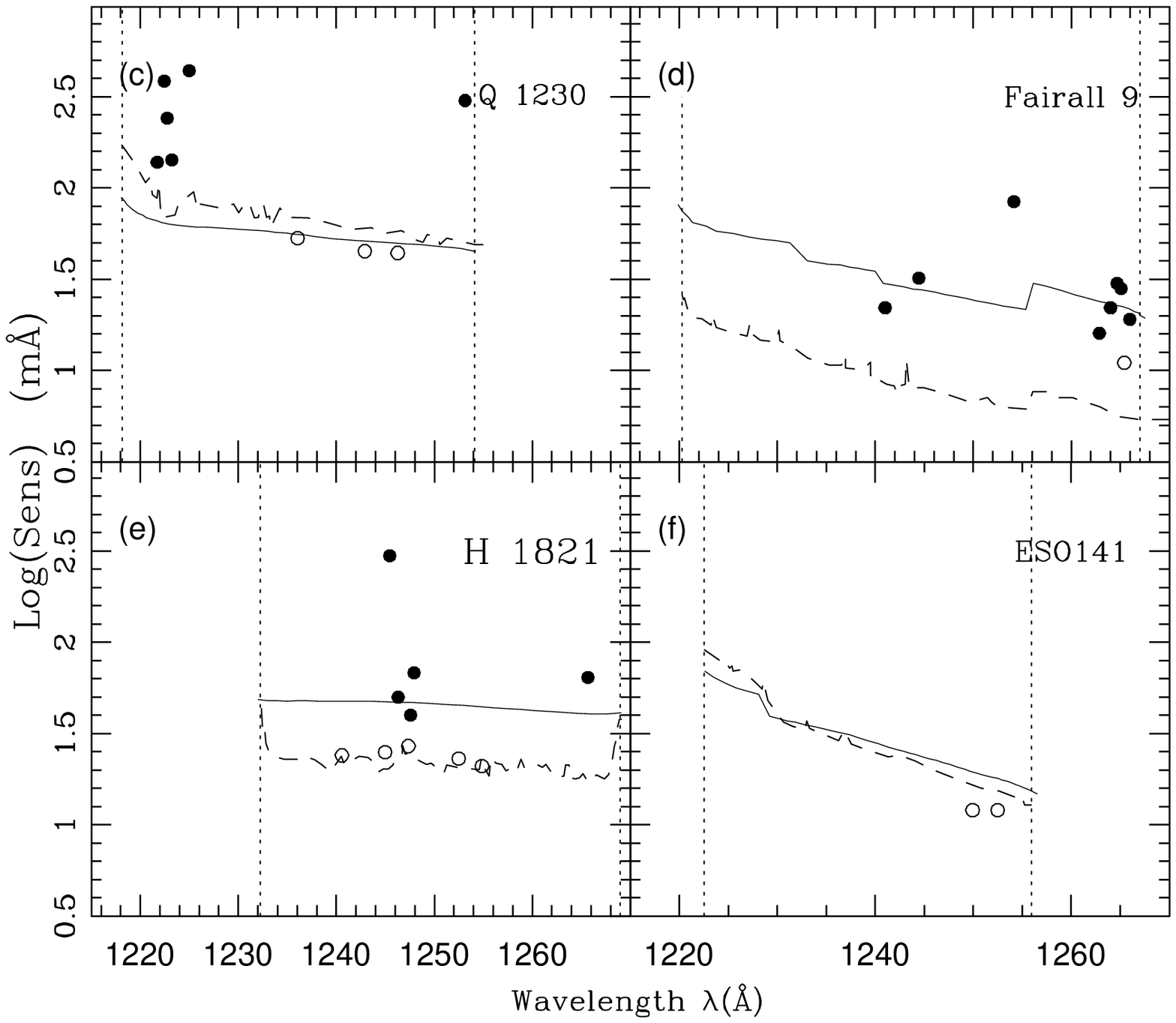}
\vspace{-2.cm}
\caption{ (Continued) }
\end{figure}

\begin{figure}[h!] 
\setcounter{figure}{1}
\centering
\epsscale{1.}
\plotone{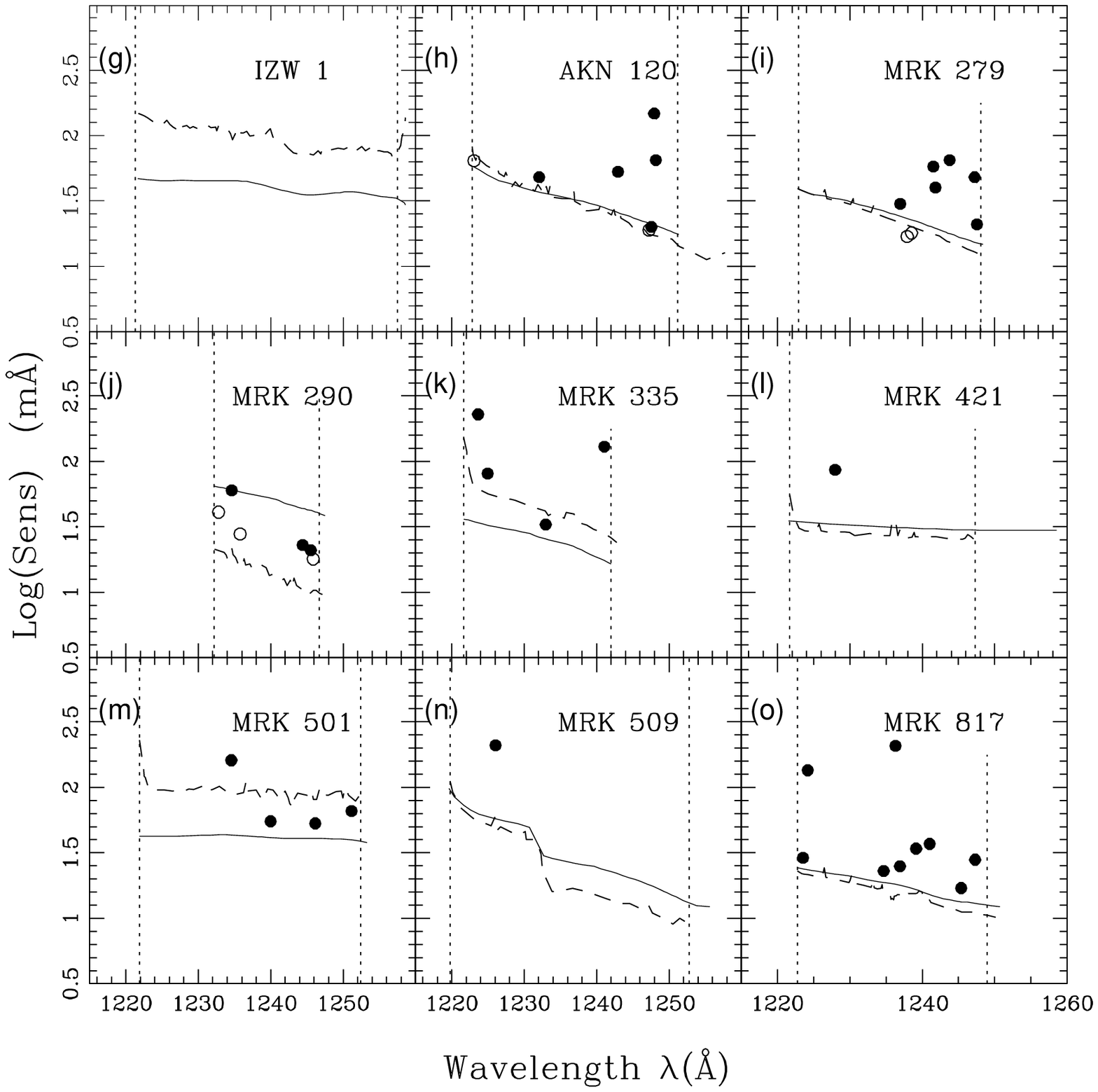}
\vspace{-1.cm}
\caption{(Continued)}
\end{figure}
The sensitivity functions published by PSSI (the dashed lines in Fig. 2)
are based on the noise in the spectrum around the placed continuum.
Clouds are plotted at their observed \lya\ wavelength, and rest EW.
Filled circles represent the PSSI assignment of $SL$ values $\ge
4-\sigma$, while clouds with 3-$\sigma\le SL < $ 4-$\sigma$ are shown
by open symbols.

It is apparent that the two sets of sensitivity functions do not
always agree.  While many are in substantial agreement (\eg, Mrk 421,
Mrk 279, Akn 120, Mrk 817, Q1230+0115, ESO 141-G55 and possibly Mrk
509), serious problems are apparent for IZW 1, Mrk 290, Mrk 501,
Fairall 9, and H1821+643.  Some sightlines appear to have clouds with
$SL \ge 4-\sigma$ and EWs less than the sensitivity function, or
clouds with $SL < 4-\sigma$ with EWs larger than the sensitivity
function (\eg, Figs. 2a, d, e, j, k, and m), which violates the
derivation given above for a 4-$\sigma$ sensitivity function.  This
may indicate a source of error which is outside the normal physical
explanations.  There may be ``grey'' areas in the data processing
which can introduce substantial errors.  For instance, unrecognized
3-$\sigma$ clouds, and possible continuum fitting errors may cause an
over-estimation of the continuum noise, and a correspondingly
under-estimated significance level of a given cloud.  There are four
cases in which cloud EWs are significantly larger (in m\AA) than the
flux-based sensitivity function at that location (3C 273, IZW1, Mrk
335, and Mrk 501) which may be explained by this logic.  But there are
a similar number in which their sensitivity functions are lower, and
this cannot be explained in the same manner.  The cause of this
variation is uncertain.  However, since the sensitivity functions
based on the flux in the continuum is a simpler relation, the author
judges that the flux-based sensitivity functions are more likely to
reliably reflect the actual sensitivity.  For this reason our standard
processing of data uses flux-based sensitivity functions.  The
significance levels of the cloud are hence re-evaluated according to
their EW relative to the sensitivity function.

\subsection{The Galaxy Catalogs}

A catalogue of galaxies is constructed along each LOS, and this is
used this to calculate ${\cal T}$.  All galaxy catalogs are taken from
the CfA Redshift catalog, maintained by J. Huchra
(\citealt{Huchra:90}, \citealt*{Huchra:95}, \citealt*{Marzke:96},
\citealt*{Grogin:98}, \citealt*{Huchra:99})\footnote{see
http://cfa-www.harvard.edu/~huchra/}.  The data which the catalog has
that is of use to our enterprise is the position (B1950), velocity
(with error), magnitude (with error), and the de Vaucouleurs T-Type.
The T-Type is negative for early-type galaxies and quasars and
positive for spirals and irregulars.  This allows one to pick out
those galaxies which are not well-described by the Tully-Fisher
relation.  The catalogs specific to each line of sight include all
galaxies in the CfA catalog which are within a cylinder $7.5 \, {\rm
h}_{75}^{-1}$ Mpc in radius, and within 15 degrees of the line of
sight, and which have a redshift range extending 750 \kms\ beyond the
spectral endpoints (see Table 1).  We discuss the completeness levels
of these catalogs in \S5.4.


The radial positions of both galaxies and clouds relative to the
observer are assumed to be that attributed on the basis of the
redshift alone; clouds and galaxies are assumed to have a vanishing
peculiar velocity (see \S5.5 for further comment on this).  Galaxy
apparent magnitudes are assumed to be blue magnitudes, though the
specific filter varies, and there is a low level of ``contamination''
with other filters when blue is not available.  The Tully-Fisher
relation (\citealt{Tully:77}, hearafter, TF) is used to determine the
circular velocity range (see \S5.1.1), and from this, the galaxy mass
as a function of halo model.  Though not all galaxies are spirals, we
can adjust for this by noting that galaxies with negative de
Vaucouleurs T-Types generally have mass-to-light ratios 2 to 4 times
greater than spirals.  We discuss this detail in \S.6.2.
 
Much depends on the reliability of the apparent magnitudes in the
assembled galaxy catalogs.  For those interested in this level of
detail, the Appendix offers a reconciliation of the CfA magnitudes
with magnitudes extracted electronically from digitized Palomar
Observatory Sky Survey plates, where we conclude that the CfA magnitudes are accurate to $\pm 0.4$ mag..


\section{The Galaxy Mass and Tidal Fields} 

The next task is to calculate the tidal field produced by the galaxies
in each catalog for each point along the lines of sight.  We must
first discuss the derivation of galaxy masses, then the tidal field.
We also inspect the results of applying this methodology to the
galaxy catalogs along the sightlines to AGN.

\subsection{Galaxy mass}
The basic procedure for deriving a galaxy's mass is as follows:
apparent magnitude and redshift are used to derive an absolute
magnitude, then an empirical relation must be exploited to estimate
the galaxy mass.  The overwhelming fraction of galaxies are of spiral
type.  For those, the TF relation is used to determine the circular
velocity of the galaxy, and hence its mass as a function of the
adopted halo model.  The TF relation expresses the luminosity as a
power of the circular velocity,
\begin{equation}
{\cal L} \propto v_c^{\beta},
\end{equation}
where $v_c$ is the maximum circular velocity of the galaxy, and the
slope $\beta$ is a function of spectral range.  
\citet{Tully:00} show that the blue-band slope, relevant to the CfA
catalogue, is $\beta_B = 2.91$.  The B-band galaxy luminosity is
influenced by recent star formation, and so is more likely to
experience larger excursions from the mean relation than, for
instance, the K-band relation.  For individual galaxies, there is a
0.55 mag dispersion relative to the calibrated $B-$band relation
\citep{Tully:00}, but only 0.4 mag in $R$ and $K^{\prime}$.  Given a
DM halo model, one can extract a galaxy mass with error.

The derived mass is a function of the halo model.  But having accurate
\emph{relative} masses of galaxies is more important at this stage
than achieving accurate absolute masses.  We consider two halo types.
Our standard model is a truncated isothermal halo, and the NFW model,
is treated as the first alternative.  We treat the two halo models in
order below.

\subsubsection{The mass of truncated isothermal spheres}

We first assume that the distribution of galaxy mass is well-modeled
by a truncated isothermal sphere.  In this case, the halo mass density
can be approximated by,
\begin{equation}
\rho(R)=\frac{\cal K}{4 \pi R^2}, .................(R \le R_t),
\end{equation}
where ${\cal K}$ is called the total mass distribution constant, and
$R_t$ is the mass--truncation radius.  Since the TF relation is based
on circular velocities for dark matter plus baryons, ${\cal K}$ is taken
to be the observed mass-distribution constant, where
\begin{equation}
{\cal K} = \frac{{v_{c}^2}}{G}.
\end{equation}

Assume that the galaxy masses are truncated at a radius such that the
density at $R_t$ is the same for all galaxies.  A glance at Eq. 27 will
show that the truncation radius $R_t \propto {\cal K}^{1/2}$.  With
the above relations (Eqs 26, 27, and 28) we may construct the following
scaling relations valid for isothermal halos:
\begin{equation}
\frac{v_c}{v_c^*} =\biggl( \frac{\cal L}{{\cal L}^*} \biggr)^{1/\beta_B} = 
\biggl(\frac{\cal K}{{\cal K}^*} \biggr)^{\frac{1}{2}} = \frac{R_t}{R_t^*} =
\biggl(\frac{{M}(R_t)}{{M}^*(R_t)}  \biggr)^{\frac{1}{3}},
\end{equation}

This halo model is standardized by requiring that an ${\cal L}^*$
galaxy has a truncation radius $R_t$ of $0.5 \, {\rm h}_{75}^{-1}$
Mpc, consistent with the observations by
\citet{Zaritsky:94} and \citet{Zaritsky:97a} of the steadiness of velocity
dispersions of satellite galaxies at large galactocentric radii around
parent field spirals.

With the use of the cataloged redshift, the absolute magnitude of the
galaxies is calculated.  The Tully-Fisher relation is used to derive
the circular velocity.  The relevant error in absolute magnitude is
the CfA magnitude error ($\pm 0.4$ mag, Appendix) combined in quadrature with
the TF error per galaxy (0.55 mag in the $B$-band), which results in a
total error of 0.68 mag.  According to \citet{Tully:00}, the maximum
circular velocity of a galaxy with absolute magnitude of $M_B$ is,
\begin{equation}
v_c = \left[158 \, exp\left(\frac{-(M_B+20.11-5
\log{h}_{75})}{7.27}\right)\right]_{-0.19 v_c}^{+0.255 v_c} \, \kms,
\end{equation}
where the 1-$\sigma$ errors in attributed total circular velocities
are shown.  The calibration of \citet{Tully:88} indicates that $M_B^*
= -20.18 + 5 \log{\h_{75}}$.  The total mass ${M}^*$ of an ${\cal
L}^*$ galaxy is then,
\begin{equation}
{M}^*=\frac{(v_c^*)^2 \, R_t}{G}
\end{equation}
where $v_c^*$ is the circular velocity of a ${\cal L}^*$ galaxy.  With
the scaling equations (Eq. 29) we see that $M \propto v_c^3$.  The
 mass of a galaxy may be expressed as,
\begin{equation}
{M}=\left[{M}^*{\left(\frac{\cal L}{{\cal L}^*}\right)^{3/\beta_B}}\right]_{-0.35 M}^{+0.54 M},
\end{equation}
where ${\cal L}/{\cal L}^*=C \,\exp({-0.4 M_B})$, and $C=8.47 \times
10^{-9}$, and $\beta_B=2.91$.  The 1-$\sigma$ range
in observed mass $M$ is then $0.65 {M}$ to $1.54 {M}$.

Note that if one sets $M_B=M_B^*=-20.18$ mag ($h=0.75$)
\citep{Tully:88}, one can solve for $v_c^*$ by substituting $M_B^*$
and using Eq. 30.  In the B-band, \be v_c^*=161.5 ~\kms.  \ee 

What evidence, you might ask, is there to support this halo model
(other than flat rotation curves)?  Each halo at large galactocentric
radius, and the gas associated with the halo, is assumed to scale as
shown in Eq. 29, indicating that at a radius which scales $r \propto
{\cal L}^{1/{\beta}}$, the gas density of each system would be the
same.  We assume that the EW of \hone\ gas found at distance $r_p$
from a galaxy should be proportional to the gas density there.  A
recent paper by \citet{Chen:01} has calculated the radius at which a
characteristic absorption of 0.3 \AA\ occurs and found that in the
K-band, $r \propto {\cal L}_K^{0.28 \pm 0.08}$ at a characteristic
radius $r_{ch} \approx 180 \, {\rm h}^{-1}$ kpc.  In the B-band, $r
\propto {\cal L}_B^{0.39 \pm 0.09}$.  The isothermal halo model
predicts that the exponent for both is the respective $\beta^{-1}$,
which, given that $\beta_K=3.51$ \citep{Cole:01}, and $\beta_B=2.91$
yields $r \propto {\cal L}^{0.285}$, and $ \propto {\cal L}^{0.344}$,
both well within the errors quoted above.  That is, the gaseous halos
observed by \citet{Chen:01}, and the scaling laws for isothermal
halos, are consistent with each other.

On the other hand, the truncated isothermal halos proposed here are
significantly more massive than those of the favored, NFW halo.  While
the mass within the distances probed in studies of the mass-luminosity
ratio \citep*[\eg.][]{NBahcall:95} is compatable with the NFW halo, so
is the mass of an isothermal halo.  It is in the realm beyond the
virial radius, which has not been effectively probed, where this extra
mass may be found.

\begin{figure}[h!] 
\centering
\epsscale{0.95}
\plotone{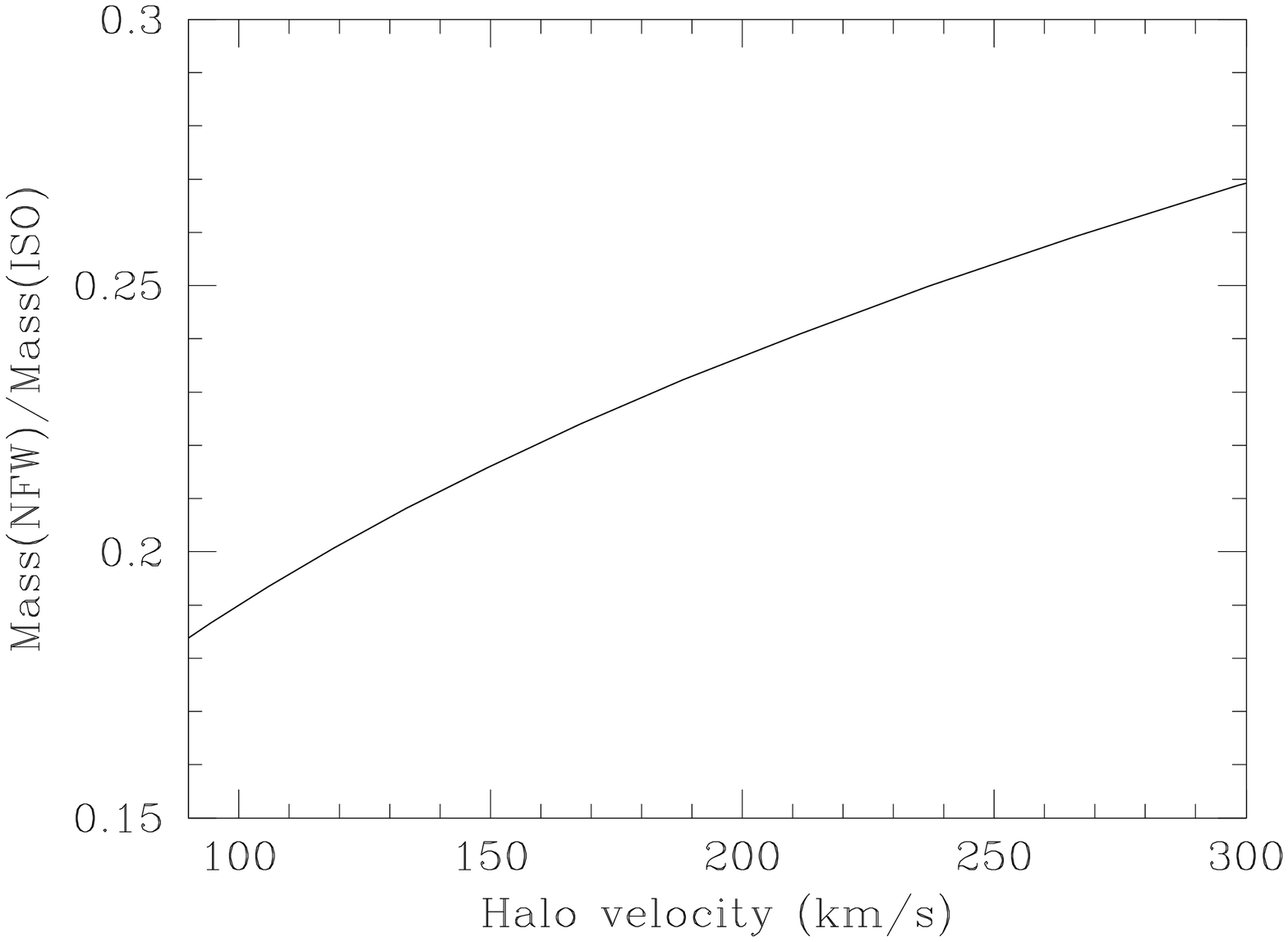} 
\caption{The ratio of the NFW virial mass to an isothermal halo
truncated at a radius scaled to that of an ${\cal L}^*$ galaxy with
$r_t=0.5 \, h_{75}^{-1}$ Mpc.  The assumption of an isothermal halo
will have the effect of attributing relatively more mass to small
halos than to large.  The skew from strict proportionality will be
less than about $10\%$ over the range of halo velocities of interest in
this paper.}
\end{figure}

\subsubsection {The mass of a NFW halo }

If halos are described as NFW, then ${\cal K} \equiv
{M}_{max}/R_{max}$, where $R_{max}$ is the distance at which the
circular velocity of the halo is a maximum.  The density is, \be
\rho(R) = \frac{\delta_c ~\rho_{crit}}{(R/R_s)(1+R/R_s)^2}, \ee (NFW,
1997), where $R_s$ is the scaling radius; $R_{max} \simeq 2.16 R_s$.
NFW halos are defined for the dark component only, and one has to
adjust for the baryon fraction.  This can be done by requiring that
enough baryons are added within $r_{max}$ to give a flat rotation
curve inward of that point.  As with the isothermal halo, the circular
velocity is derived from the apparent magnitude and the redshift
(refer to Eq. 30).  The concentration parameter $c$ is defined as the
ratio of the virial radius $R_{200}$ and the scaling radius $R_s$.

By fitting the relation in Fig. 9 of NFW (1996) we find,
\begin{equation}
\frac{v_{max}}{v_{200}} = 2.44 v_{max}^{-0.110}.
\end{equation}
Figure 10 in NFW (1996) gives,
\begin{equation}
R_{max}=1.5 \times 10^{-2} \, v_{max}^{1.49}.
\end{equation}
From Eq. 3 of NFW (1997)  we find a reasonable fit  yielding,
\begin{equation}
\frac{v_{max}}{v_{200}} = \frac{c^{1/2}}{(2.16
\left\{ln(1+c)-c/(1+c)\right\})^{1/2}} = 1.27 \, c^{0.28}.
\end{equation}
Equating Eqs. 39 and 41 yields,
\begin{equation}
c=132.3 \, v_{max}^{-0.393.}
\end{equation}
Next we find,
\begin{equation}
\frac{M_{200}}{M_{max}} = \left(\frac{v_{200}}{v_{max}}\right)^2 \,
\frac{R_{200}}{R_{max}} = 1.20 \, c^{0.44}.
\end{equation}
Using these equations we can construct scaling relations analogous to
those of Eq. 29,

\begin{equation}
\frac{v_c}{v_c^*} =\biggl( \frac{\cal L}{{\cal L}^*}
\biggr)^{1/\beta} = \biggl(\frac{R_{max}}{R_{max}^*}
\biggr)^{0.67} =\biggl(\frac{R_{200}}{R_{200}^*} \biggr)^{0.76} =
\biggl(\frac{{M}_{200}}{{M}_{200}^*} \biggr)^{0.30}.
\end{equation}
The errors propagate in the same way as with the isothermal halo
though the virial mass is proportional to $v_{max}^{3.32}$, rather
than $v_{max}^{3.0}$ for the isothermal halo.  This accounts for the
rising trend in the ratio of masses of NFW and isothermal halos of the
same $v_{max}$ (see Fig. 3).

How does this halo accord with the characteristic radius for
absorption observed by \citet{Chen:01}?  We would expect that it
should scale with the virial radius.  From the scaling relations
above, $R_{200} \propto {\cal L}^{1/(0.76 \beta)}$, yielding
$R_{200}(K) \propto {\cal L}_K^{0.375}$, and $R_{200}(B) \propto {\cal
L}_B^{0.452}$, which are supposed to compare with exponents $0.28 \pm
0.08$, and $0.39 \pm 0.09$, respectively from \citet{Chen:01}.  The
predicted $K$-band relation is outside the error bars, while the
$B$-band relation is within the 1-$\sigma$ error bars, but near the
edge.  If instead one were to require the characteristic radius to
scale with $_{max}$, then the lack of concordance is exascerbated,
giving exponents 0.43 and 0.52 for $K$ and $B$-bands, respectively,
both well-outside 1-$\sigma$ error bars of \citet{Chen:01}.

The NFW halo, however, is defined by the dark component only.  By
adding baryons until a flat rotation curve is maintained out to
$r_{max}$, one finds that the circular velocity is increased by from
$\sim 12\%$, for small clouds, to $\sim 8\%$ for large galaxies.  The
NFW halos are corrected for their baryonic part when comparing the
masses of NFW and isothermal halos.

\subsection{Error in Attributed Tidal Fields}

We have seen in general terms how the tidal field is calculated along
the LOS (\S3.1).  If we assign no error for the distance $R(i)$, the
fractional error in individual tidal fields is the same as that in the
attribution of mass to a galaxy on the basis of its redshift and
apparent magnitude $m_B$ -- that is, on order 50\%.  The error in
$R(i)$ will almost entirely result from positional errors along the
line of sight (see \S3.1).  The fractional error in the attributed
tide produced by a number of different galaxies will generally be
significantly less than from a single galaxy because the individual
errors are independent.  However, since the masses and tides vary,
there is no simple way to calculate the total error except by a
specific convolving with redshift and mass errors of the effects of
each galaxy.  This level of sophistication is not pursued here,
however, we will discuss the errors resulting from cloud peculiar
velocities and redshift errors (\S5.5).

Another source of error is the volume from which the galaxy catalogs
are drawn.  Let us consider the tidal field of an ${\cal L}^*$ galaxy
at the edge of the 7.5 Mpc cylinder at the same redshift of a cloud (a
minimum maximum distance, so to speak).  For an NFW halo the vector
tide at 7.5 Mpc is ${\cal T}_R\simeq 0.00195$, while the isothermal
halo is $\sim 5$ times greater, ${\cal T}_R \simeq 0.0098$.  This tide
would produce a scalar tide of ${\cal T}=6.5 \times 10^{-4}$, and $3.3
\times 10^{-3}$ for the NFW and isothermal halos, respectively.  Thus
our estimated relative tidal fields may begin to produce serious
errors at respective tides of this order.  We shall see (\S5.4,
Fig. 5) that only a small fraction of galaxy mass is contained in
galaxies brighter than ${\cal L}^*$, so the magnitude of the error is
generally small.

\begin{figure}[h!] 
\setcounter{figure}{3}
\centering
\epsscale{1.0}
\plotone{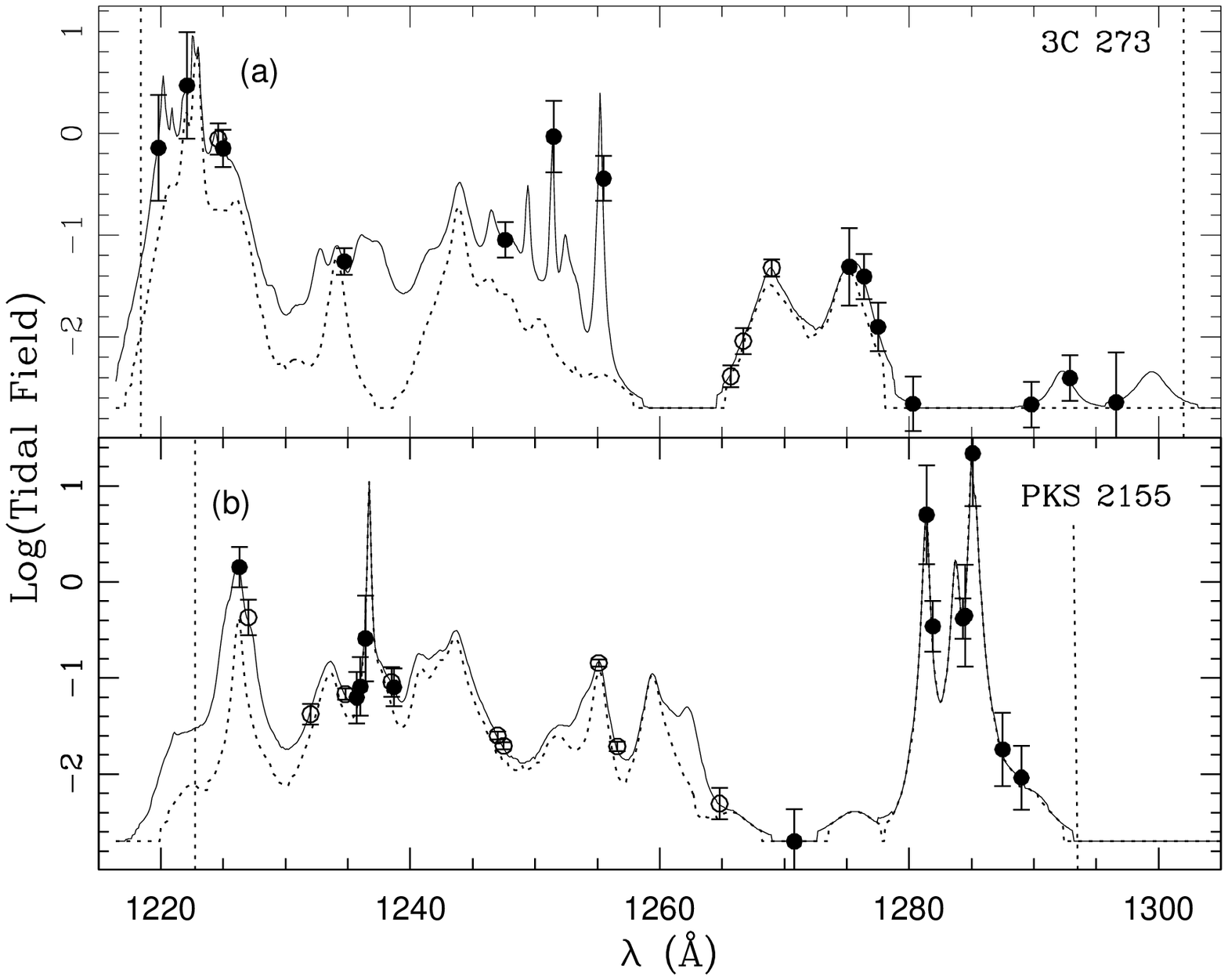}
\vspace{-1.3cm}
\caption{Tidal fields along the lines of sight to target AGNs.  Panels
$a$ and $b$ show 3c273 and PKS 2155-304.  The tides are expressed in
units of the tidal field parameter ${\cal T}$ (\S3.1), which is
convenient for constraining average cloud density for stability
against disruption (see Eq. 19).  The solid lines represent all galaxy
data, while the dotted lines represent the effects of only those
galaxies brighter than $M_B=-19.8$ mag, the absolute magnitude of a
$m_B=16$ galaxy at $z=0.036$ (see \S5.2).  The highly significant
(4-$\sigma$) absorption systems, in terms of the author's sensitivity
functions (solid lines, Fig. 2 $a$-$o$, are denoted by the solid
circle), while the 3-$\sigma$ clouds are represented by open symbols.
The cloud EWs are denoted by the error bars, with length proportional
to $\log{\cal W}$.  The vertical dotted lines show the limits of the
useful portion of the absorption system surveys (see \S3.1 and Table
1).}
\end{figure}

\subsection{Sightline Characteristics}

As mentioned above (\S5.2), the LOS component of the tidal field is
calculated assuming a pure Hubble flow of galaxies, while those normal
to the LOS are calculated using the appropriate angular diameter
distance relation.  Figure 4 shows the calculated tidal fields (solid
line) as a function of position along the 15 LOS.  Cloud positions are
noted by the circles (filled, $SL \ge 4$-$\sigma$, open, $SL <
4$-$\sigma$) with vertical bars, whose lengths are proportional to
$\log{\cal W}$.  Note the similarity of the tidal fields in the
sightlines to 3C 273 and Q1230+0155, which both pass through the Virgo
cluster.  Some stretches of space in the various lines of sight have
no cataloged galaxies within the search criteria, and so have a
formally zero tide.  A value of 0.002 is added to all tides so that
the line would appear on the plots.  The significance of the dotted
lines which ``shadow'' the tidal field traces is discussed in the next
subsection.

\begin{figure}[h!] 
\setcounter{figure}{3}
\centering
\epsscale{1.0}
\plotone{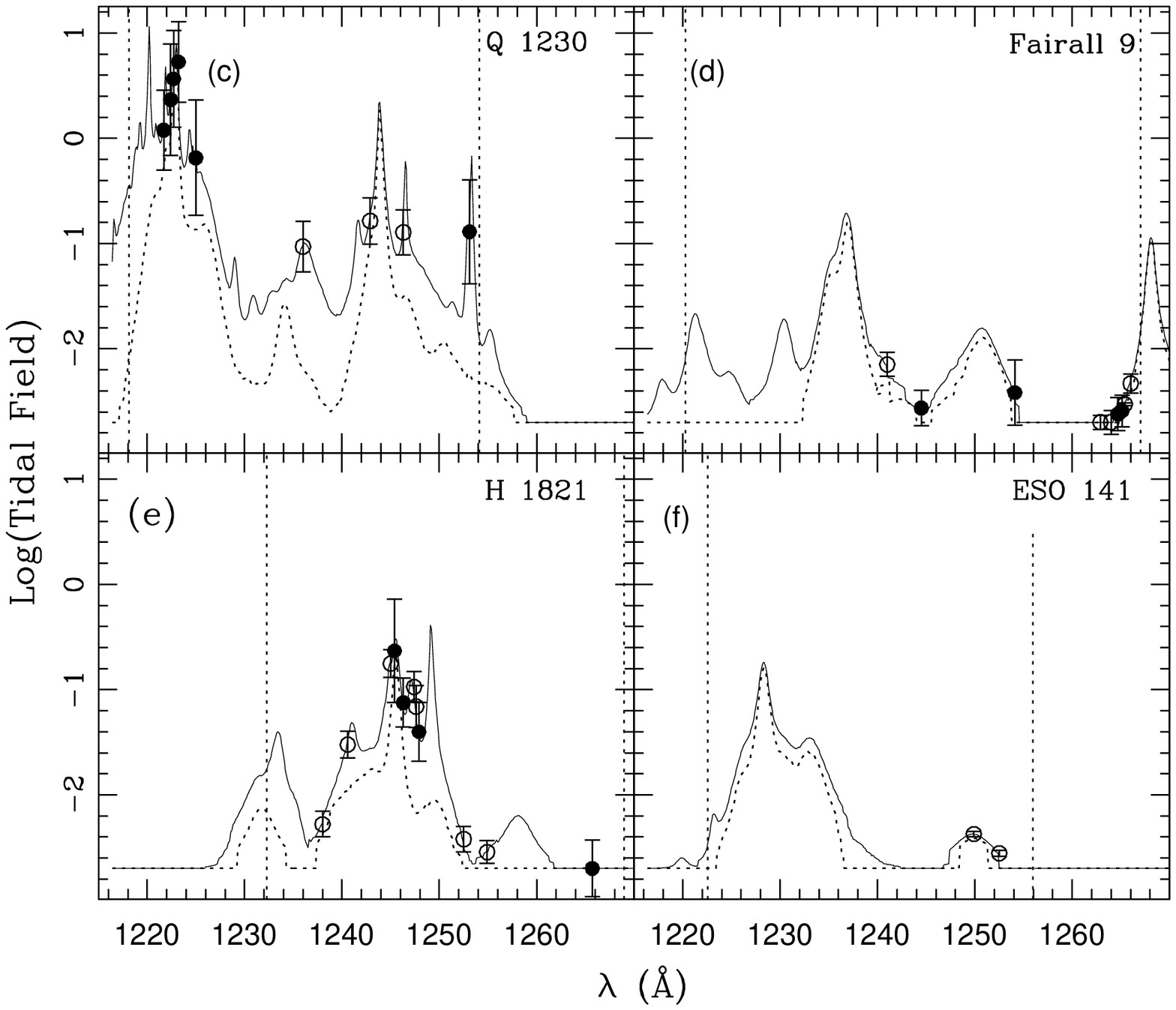}
\vspace{-1.0cm}
\caption{(Continued)}
\end{figure}

\begin{figure}[h!] 
\setcounter{figure}{3}
\centering
\epsscale{1.0}
\plotone{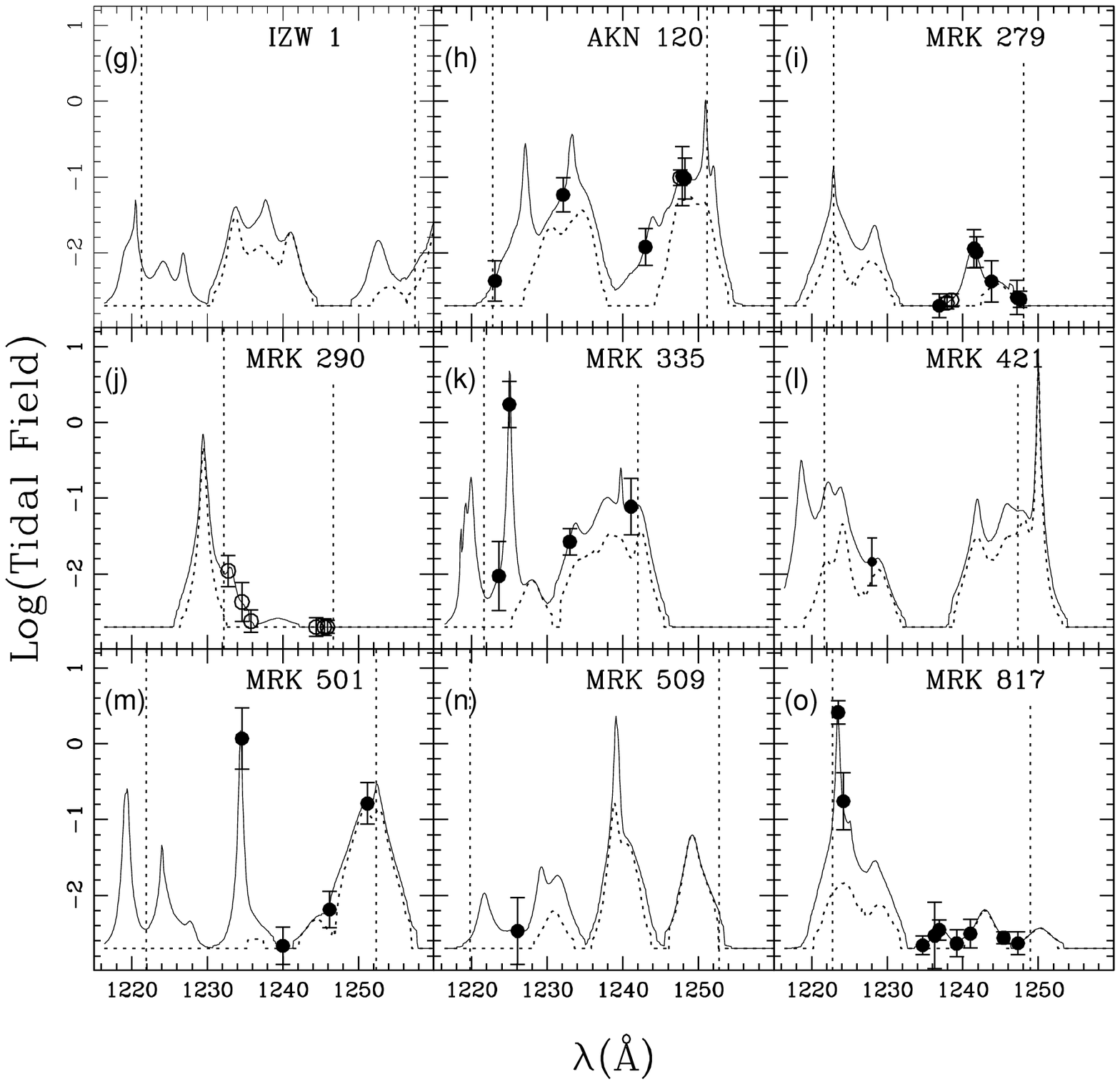}
\vspace{-1.1cm}
\caption{Continued)}
\end{figure}

\subsection{Effect of Lack of Completeness }

Though galaxy detections are surface brightness selected, the decision
to extract a redshift and place an object in a galaxy catalog are
often determined by the total apparent magnitude.  The CfA catalog is
generally based on Zwicky magnitudes and have a completeness limit of
order $15.5 \le m \le 16.0$ mag, but the catalog accepts galaxy data
from a wide range of sources.  To get an idea of how much mass is
unrepresented in the catalogs, a fractional cumulative mass function
is calculated.  Figure 5 shows the cumulative fractional mass function
for the truncated isothermal halo in terms of the absolute magnitude
of the galaxy with a faint-end slope $\alpha=-1.2$ (solid line).  It
is produced by the following integral, \be f_M(<v_c)=
\frac{\int_{v_i}^{v_c} M(v_c) \phi(v_c) dv_c}{\int_{v_i}^{v_f} M(v_c)
\phi(v_c) dv_c}, \ee where $v_i=37$ \kms and $v_f=375$ \kms, which
encompasses the overwhelming majority of galaxian mass, and $\phi(v)$
is the B-band luminosity function transformed to a velocity function
by use of the TF relation.  With the use of Eq. 30 we express halo
velocity in terms of absolute magnitude.  Note that $\sim 35\%$ of the
galaxy mass is in the range $M_B \la -19.8$ mag.  A galaxy with $M_B =
-19.8$ is 0.74 ${\cal L}^*$ ($v_c\sim 146$ \kms), and has an apparent
magnitude of 16 mag at $z\simeq 0.036$.

However, it may be reasonable to suggest that the sub-halos within a
given halo are characteristic of the system, and not an excess over
the halo attributed to the luminosity of the central galaxy.  The NFW
halo is the profile for an individual halo.  Studies of ``secondary''
infall \citep{Zaroubi:93, Avila-Reese:99, Klypin:99} suggest that
sub-halos are not included in the mass distribution of the standard
NFW halo.  The truncated isothermal halo, however, describes a
\emph{system} extending far beyond the galaxy which has formed at its
center.  In this case, the sub-halos are interpreted as an aspect of
the system referred to by the mass distribution constant ${\cal K}$
(see Eq. 28).  So in the case of the isothermal halo, we count
sub-halos twice unless a factor representing the fraction of halos
with a given halo velocity which are not sub-halos is inserted into
both upper and lower integrals of Eq. 41.  The fraction of halos of
velocity $v_c$ which are isolated has been estimated by
\citet{Avila-Reese:99}, and at low $v_c$ it appears to follow the
relation $f_{iso}=1.36 (v_c/225) -0.23$ for halos smaller than 158
\kms\ ($0.7 \times 225$).  The results of applying $f_{iso}$ to the
the numerator and denominator of Eq. 41 is the dotted line in Fig. 5.
While with the former $M_B=-19.8$ mag occurs at $f_M \simeq 0.65$,
with the latter it occurs at $f_M=0.44$.  The implication of this is
that the sub-halos of the brighter galaxies are counted already, so
less is missed.  On the other hand, it now appears possible that we
may have over-estimated the tidal field at very low-redshift by
observing their sub-halos, and thereby effectively counting them
twice.  There is thus an inevitable tendency to find more mass at low
redshift.  Above $z \sim 0.036$, about half of the mass in galaxies is
missed for catalogs with a uniform magnitude limit $m_B=16$.

Four spectra extend above $z=0.036$, 3C 273, PKS 2155-304, Fairall 9,
and H1821+643 with a total of 21 clouds.  To test for the effects of
incompleteness one could calculate the equivalent width distribution
with, and without, the data from $z\ge 0.036$.  If there is no
significant difference, then one might say that completeness issues
are not debilitory to the conclusions of the paper, for if no
completeness effects found by the removal of data with $z \ge 0.036$
then it is unlikely to do so at lower redshifts.  However, we should
expect that clouds at higher redshift may have calculated ambient
tidal fields a factor of $\simeq 2$ low, hence one runs the risk of
placing what may be GDS clouds in void catalogs.

A sense of the potential effects of incompleteness can be seen by
placing a uniform limit on the luminosity of galaxies used to
calculate tides.  In Fig. 4 (panels a-o) the dotted line represents
the tidal field when galaxies with $M_B$ fainter than $-19.8$ mag are
removed from consideration.  Generally the discrepancy between the
total and the contribution from only the brighter galaxies declines
with increased redshift as fewer and fewer low luminosity galaxies
make the catalog (a good example of this is panel $b$).  This is
characteristic of fields of roughly uniform apparent magnitude limits.
On the other hand, the sightline to 3C 273 (panel $a$) has regions at
high-redshift where removing galaxies less luminous than $M_B=-19.8$
has a strong effect on the calculated tidal field.  This is because
some effort has been expended to get the redshifts and magnitudes of
galaxies along the line of sight \citep{Morris:93}, with a claimed
completeness to $m_B=19.0$, so that one ought to be able to detect a
galaxy of $M_B=-17.0$ at $z=0.065$.

\begin{figure}[h!] 
\centering
\epsscale{0.9}
\plotone{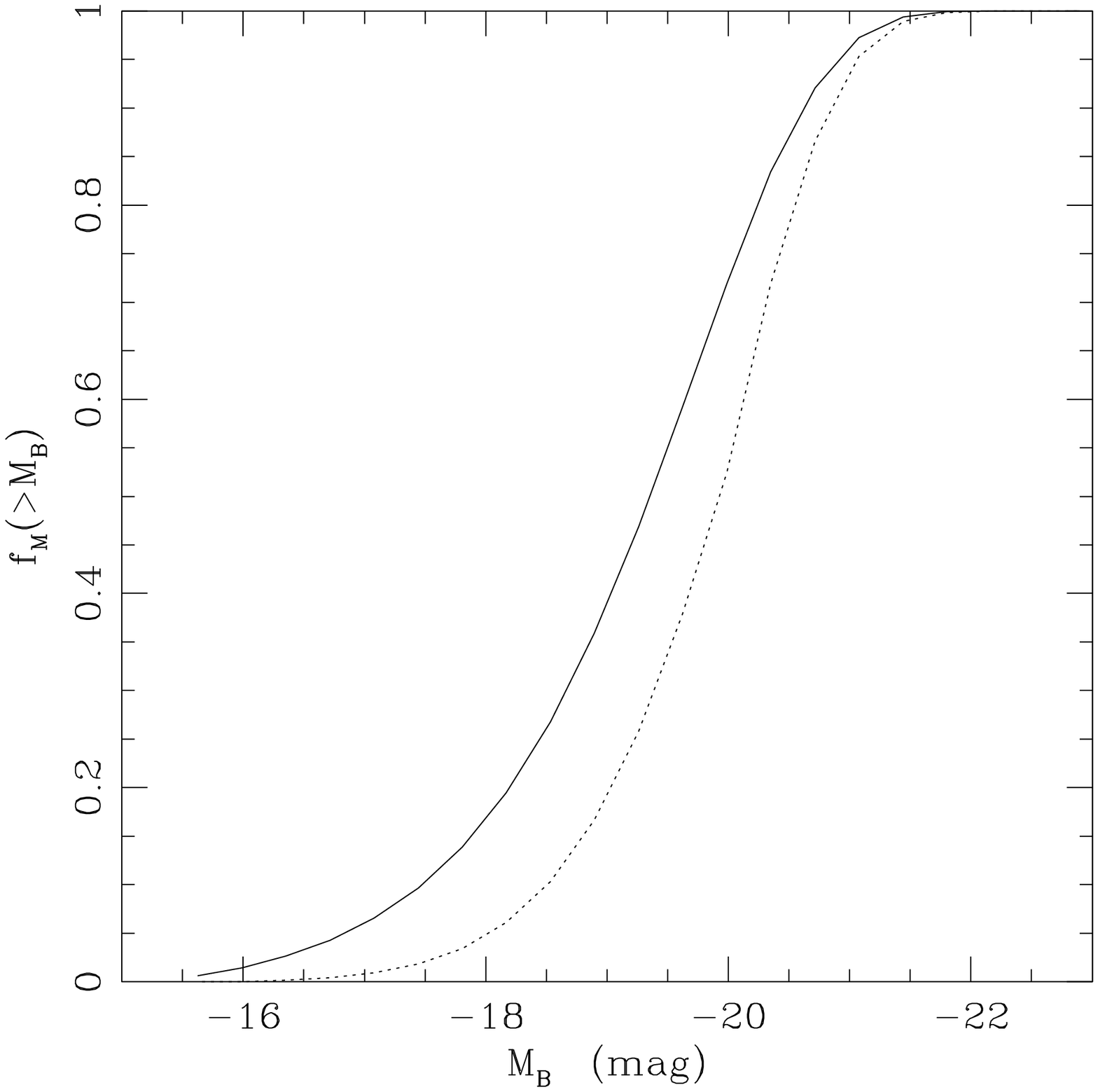}
\vspace{-.0cm}
\caption{The normalized mass function as extracted from the Schechter
Luminosity function (using $\alpha=-1.2$) and the TF relation (solid
line), both in the B-band.  The dotted line is the mass function
corrected for sub-halos (see \S5.4). }
\end{figure}

\subsection{Attributed Positional Errors}

For clouds associated with galaxies, there are two major sources of
error in the assignment of a value for ${\cal T}$ at the cloud's
location, galaxy redshift errors, and peculiar velocities,
particularly of the cloud.  We discuss these in order below.

In this paper the main concern is with void clouds, which are only
slightly effected by galaxy redshift errors since typical redshift
errors, on order 20-30 \kms, imply modest positional errors relative
to void clouds ($\sim 260$ to 400 \kpc).

Cloud peculiar velocity errors stem from the ``triple-value
ambiguity'' noted by \citet{tully:84} in which a given LOS velocity
may occur at three different distances when the object is infalling
toward a larger mass.  If clouds are responding to the peculiar
gravity of a galaxy, they may acquire maximum peculiar velocities not
much greater than $v_p \approx v_c = \sqrt{G {\cal K}}$ (see Eq. 28)
when the clouds are close to the galaxy\footnote{If peculiar
velocities are a measure of the galaxy potential, then with the
truncated isothermal halo, the peculiar velocity will be of order $v_p
\simeq 2^{1/2} v_c \sqrt{R_t/R}$.}.  A peculiar velocity of this
order, under a strict Hubble flow, results in an attributed distance,
$ R_{\sigma} = \sqrt{G {\cal K}}/H_0$.  The average over-density
within this distance is, \be {\bar \rho}(R_{\sigma}) = 2 \rho_{crit}
\frac{G {\cal K}R_t}{H_0^2 R_{\sigma}^3} = 2 \rho_{crit}
\left(\frac{R_t}{R_{\sigma}}\right).\ee For the Galaxy (assume
$v_c=220 $ \kms), the truncation radius $R_t$ is scaled up from
$R_t^*=500 ~\h_{75}^{-1}$ kpc to 684 kpc.  The over-density within
radius $R_{\sigma}$ is ${\bar \rho} \simeq 0.46 \rho_{crit}$, where in
this case, $R_{\sigma} \simeq 3.0 \, \h_{75}^{-1}$ Mpc.  The median
position angle (the angle between the LOS and the vector from cloud to
galaxy) for clouds distributed isotropically about a galaxy is 60\deg,
leading to the conclusion that the average LOS velocity will be
$\sim$1/2 of $v_p$.  Therefore, low cloud velocities relative to a
galaxy suggest either that the cloud is distant, or that its velocity
has dissipated against the gaseous halo surrounding the galaxy.

It is also possible that clouds which are actually distant (say, 2 to
3 Mpc from a large galaxy), but have a small impact parameter, may
appear at very nearly the same velocity as the galaxy; a slight
peculiar velocity may balance the Hubble flow if the cloud is near the
turnaround radius.  Thus, clouds which are within ${R_\sigma}\simeq
\sqrt{G {\cal K}}/H_0$ of a galaxy may not have their positions
accurately predicted by the redshift due to the peculiar velocity.
Such a huge positional error is not likely for the average cloud in
GDS since it would require a low-probability projection of the cloud
at large radius $R\approx3$ Mpc (where the cloud density is low) to a
small projected radius $R_p$.  However, again, this kind of error
cannot substantially effect the position of a typical void cloud, nor
its attributed ${\cal T}$.

Small groups of galaxies may, by their virial
velocities, introduce error into their attributed positions.  We have
seen that voids may be considered to begin at distances of $\approx
2$ Mpc from a single $\sim {\cal L}^*$ galaxy (\S2.1), but that
the distance at which a cloud is at the same degree of isolation from a
group will be larger.  \citet{NBahcall:93} show that the
one--dimensional velocity dispersion in groups and clusters can be
approximated by
\begin{equation}
\sigma_{LOS} \approx (90 \pm 20) \,N_g^{1/2} \, \kms,
\end{equation}
where $N_g$ is the number of galaxies in the group.  For $N_g=10$, we
have $\sigma_{LOS}=284 ~\kms$, or a spread of $\delta r \sim 1.9
h_{75}^{-1}$ Mpc on either side of the center.  But the scaling
relations (Eq. 29) show that the radius at which a given ${\cal
T}$would occur is $\sim 10^{1/3}$ times farther than for a single
galaxy.  If we consider 2 Mpc as a plausible minimum distance from a
major galaxy for a void to begin (at 2 Mpc from an ${\cal L}^*$ galaxy
the tide ${\cal T}_R=0.24$), then this translates to a distance $\sim
4.3$ Mpc for the group.  If the group is at the same redshift as the
cloud, most of the group galaxies would have attributed radial
positions within an angle of $\la $24\deg\ of the center of mass,
as seen from the cloud's position, producing a distance variation of
less than $\sim$9\%.  The tidal field at a distance of $4.3 \,
h_{75}^{-1}$ Mpc from this group would be about 86\% of that
calculated assuming its galaxies were located at the group center.
This error is significantly less than the error in an individual
galaxy.  Further, since the real apparent magnitudes of the group
should differ randomly, the error in the tidal field at the cloud
should have a lower value from apparent magnitude error than a cloud
subject to the influences of just a few.  An error of this magnitude
is insignificant in the present analysis.

\subsection{Tides and Cloud Statistics}

Sightline analysis results in a value for tidal field ${\cal
T}(\lambda)$ along each LOS (Fig. 4).  The value of the specified
tidal field limit separates the cloud population into two catalogs --
those defined by ${\cal T}$ as an upper limit ${\cal T}^U$, and those
with ${\cal T}$ as a lower limit ${\cal T}^L$.  Histograms of clouds
for various ${\cal T}$ are presented in Fig. 6, showing the former,
${\cal C}({\cal T}^U)$, as dotted lines, and the latter, ${\cal
C}({\cal T}^L)$, as solid lines.  In these plots, what is referred to
as ``void'' is relative to the current choice of ${\cal T}^U$.  Thus,
``void'' is relative to the stated tidal field limits.  If a
corresponding physical distinction between void and GDS clouds can be
found, then ``void'' may take on a more absolute meaning.  Note that
the void histograms have a most probable Doppler parameter of $b=30$
to 35 \kms\ over the whole range of ${\cal T}^U$.  The high-velocity
wing of the distribution suggests the presence in voids of more
massive clouds, while the cutoff at low $b$ gives the distribution a
large skewness, with values ranging from 0.87 to 1.06 in voids $0.01
\la {\cal T}_{lim} \la 0.1$.  For clouds defined by a lower bound
${\cal T}^L$, as one increases ${\cal T}^L$, the flat distribution
apparent for lower values of ${\cal T}^L$ gives way to a peaked
histogram and most probable Doppler parameter of $b\simeq 60$ \kms\
for ${\cal T} \geq 0.3$.  The skewness at ${\cal T}^L=0.3$ is -0.46.

\begin{figure}[h!] 
\centering
\epsscale{0.9}
\plotone{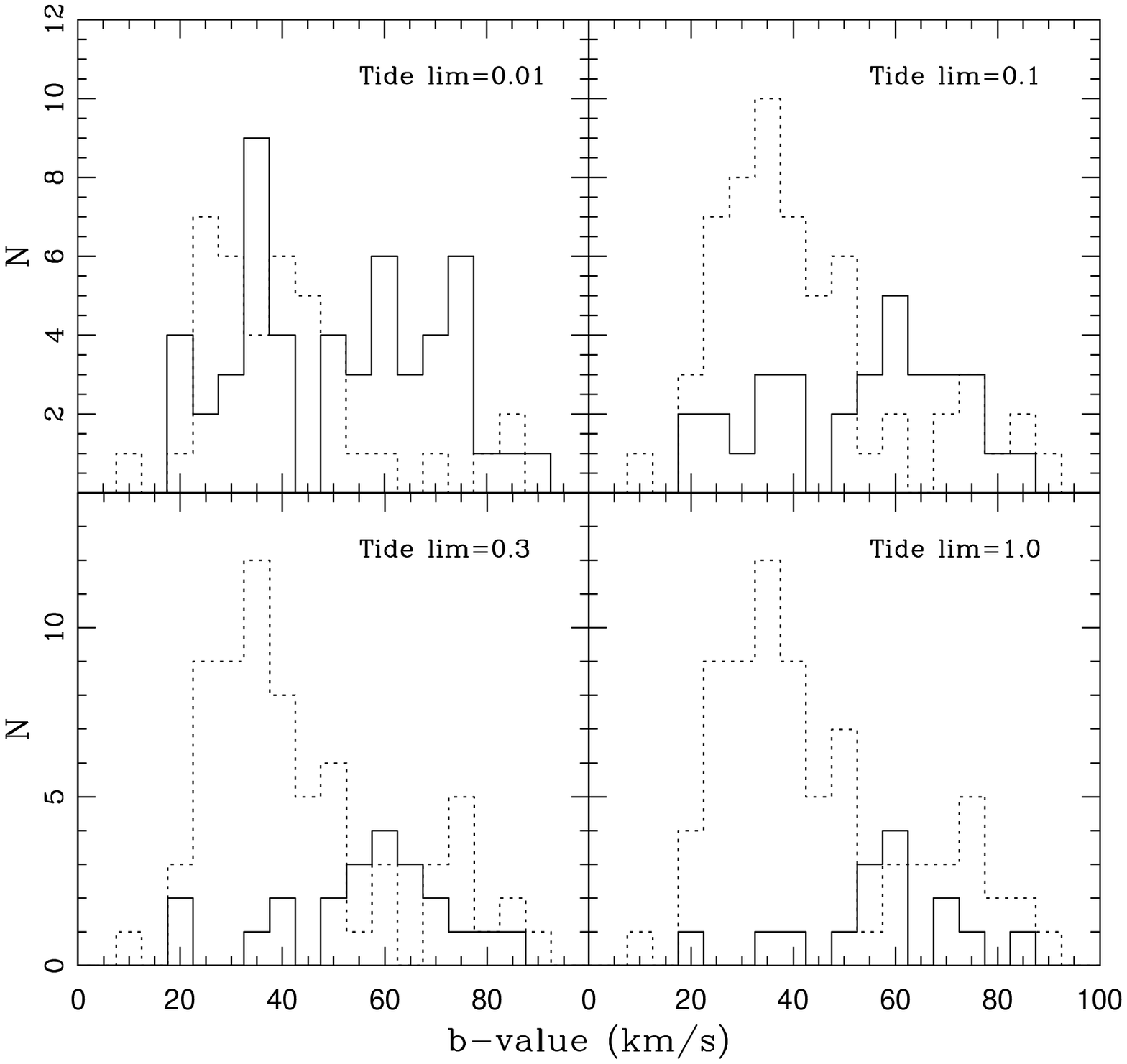}
\caption{Histograms of void (dotted) \vs non-void (solid) clouds as a
function of cloud Doppler parameter for various ${\cal T}_{lim}$.
``Void'' is defined by the value of the current tidal field upper
limit ${\cal T}^U$. Notice that the shape of the void cloud
distribution remains much the same (though its size necessarily gets
larger) as ${\cal T}^U$ increases until ${\cal T}_{lim} \simeq 0.3$.
}
\end{figure}

In Fig. 7$a$, the fractional redshift coverage as a function of tidal
field upper limit has been equated to a fractional volume which has
${\cal T} < {\cal T}^U$.  Notice that 90\% of the universe has ${\cal
T} \la 0.16$ ($\log{\cal T} \la 0.8$).  Fig. 7$b$ shows the
fractional redshift coverage as a function of EW

It is of interest to attempt to determine the point ${\cal T}^V$ which
actually separates void from GDS space.  Figure 8 shows a differential
histogram extracted from Figs. 6, which shows the number and b-value
of clouds added to the void population in the stated interval.  Panel
8$a$ shows a distribution quite similar to the overall void
population; fig. 8$b$ shows a relatively flat distribution; Fig. 8$c$
shows a distribution with characteristics of the GDS population.  This
implies that the break between void and GDS environments is at $0.1
\le {\cal T} \le 0.3$, which brackets ${\cal T}=0.16$, the tide at
which 90\% of the universe has a lower tide.  Recall that 0.9 is the
filling factor for the unshocked population (\S1).  Thus we
provisionally assign $\log{\cal T}^V = -0.8 \pm 0.2$ as the contour
dividing void and GDS space for ISOT halos, bracketing a void filling
factor $0.87 \le f_v \le 0.84$.

\begin{figure}[h!] 
\centering
\epsscale{1.}
\plotone{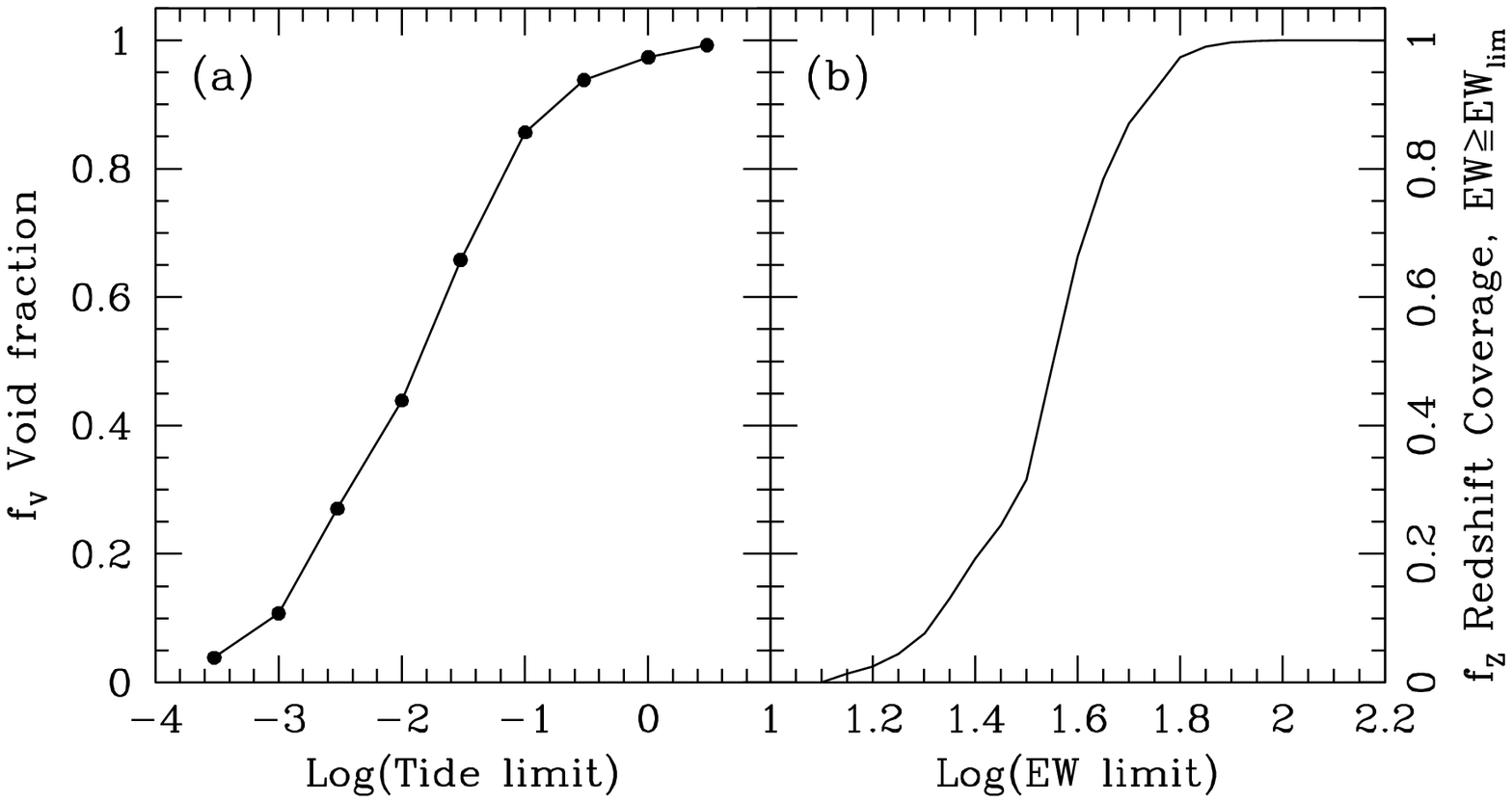}
\caption{ Panel (a): The fraction of redshift space with ${\cal T} \le
{\cal T}_{lim}$ plotted against $\log{{\cal T}_{lim}}$.  This fraction
can be considered to be a fair approximation of the volume filling
factor for voids defined by ${\cal T} \le {\cal T}_{lim}$.  Note that
tides $\log{\cal T} \la -2.5$ may seriously underestimate the
relative tidal field, as ${\cal L}^*$ galaxies farther than 7.5 Mpc may
produce scalar tides of this order (see \S5.2).  Panel (b): The
fraction of redshift space with the sensitivity function (in m\AA)
less than ${\cal W}_{lim}$ as a function of $\log{\cal W}$ (in m\AA).}
\end{figure}
\begin{figure}[h!] 
\centering
\epsscale{0.95}
\plotone{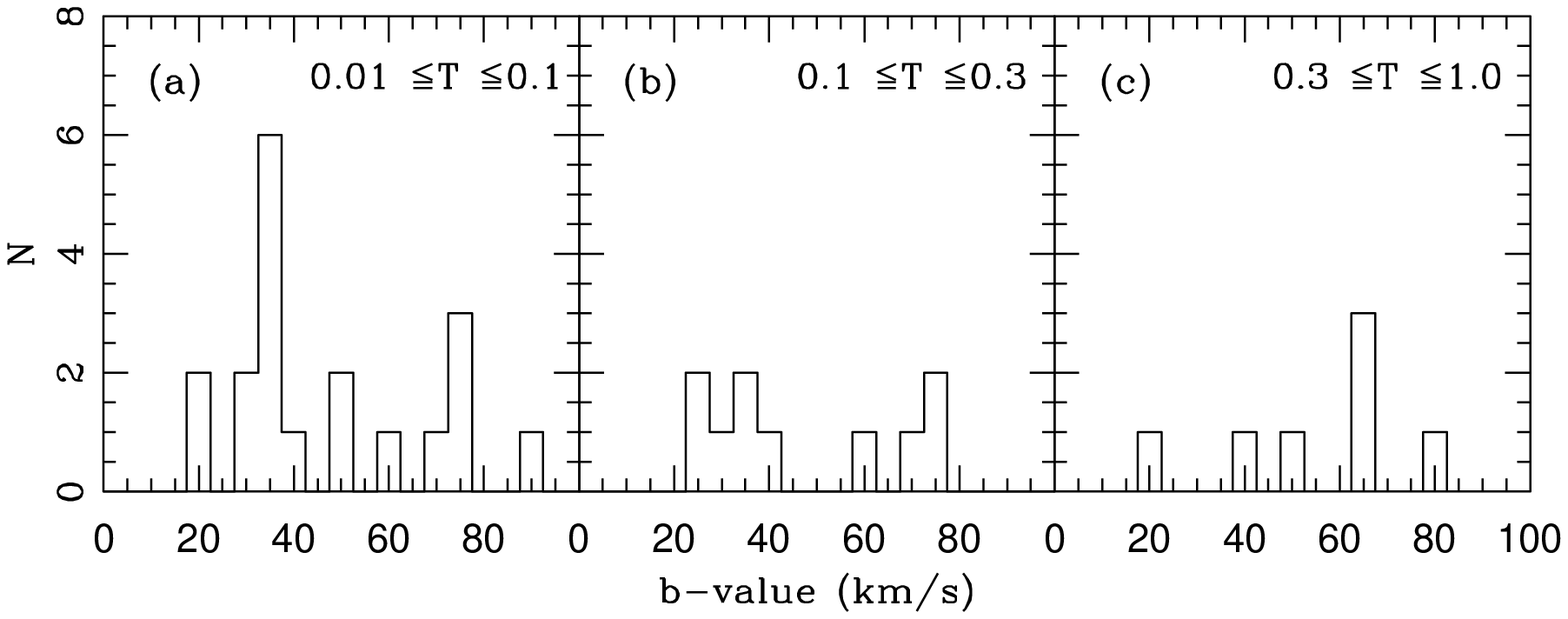}
\caption{The differential histograms from Fig. 8; histograms of clouds
added to ${\cal C}({\cal T}^U)$ for $a$ ${\cal T}^U=0.01$ to 0.1, $b$
${\cal T}^U=0.1$ to 0.3, and $c$ ${\cal T}^U=0.3$ to 1.0.  The clouds
added in panel (a) appear to have roughly the same distribution as is
found for ${\cal T}^U=0.01$ in panel (a) of Fig. 8.  Panel $b$ is
distinctly flatter indicating that clouds which are non-void may be
beginning to enter the population.  Panel $c$ begins to show the
distribution characteristic of non-void clouds, with a most likely
Doppler parameter near 60 \kms.  This trend shows that the tidal field
level at which the character of added void clouds experiences rapid
change is at $0.1 \la {\cal T} \la 0.3$.  }
\end{figure}

\section{The Equivalent Width Distribution Functions}
In \S3 it was shown that the likely effect of tides on a population of
uniform clouds is to disrupt those with density $\rho \la 2 {\cal T}
\rho_{crit}$ (Eq. 19).  However, clouds bound by dark matter are not
expected to be of uniform density (see \S2).  Because of this, it is
reasonable to suppose that the effect of tides on the distribution of
absorber EWs may have its greatest observable effect on the slopes of
the EWDF.  

We have seen that the redshift coverage at low sensitivity is restricted.  The EWDF removes this observational bias.  The EWDF of clouds of a catalog
${\cal C}$ is calculated as the redshift frequency of clouds with
rest-equivalent width greater than ${\cal W}$, directly for each
equivalent width limit ${\cal W}$ by noting the number of clouds of
any equivalent width greater than ${\cal W}$ within the ranges of
wavelengths for which the sensitivity functions have values less than
${\cal W}$.  This is then divided by the summed redshift range for
which the sensitivity ${\cal S}$ (in m\AA) is smaller than ${\cal W}$
(Fig. 7b).  Thus, the cumulative EWDF is,
\begin{equation}
f({\cal W}) = \frac{d{\cal N}(\ge {\cal W})}{dz} = \frac{ \sum_{\cal W}^{\infty}n_{cl}({\cal W}_{cl} \ge
{\cal W} \ge {\cal S})}{ \sum_{\cal W}^{\infty}\Delta z({\cal W} \ge {\cal S})},
\end{equation}
where it is read, ``$n_{cl}$ such that ${\cal W}_{cl} \ge{\cal W} \ge
{\cal S}$'', and ``$\Delta z$ such that ${\cal W} \ge {\cal S}$''.
Note that while this distribution is ostensively cumulative, it is not
impossible for the distribution to decline at low EW since there may
be conditions in which the number of clouds in the redshift interval
with very low sensitivity ${\cal S}$ is inordinately small.  So while
this distribution is cumulative, its calculation at each $\lambda$ is
to some extent independent of its neighboring bins.

Results are presented below.

\begin{figure}[h!] 
\centering
\epsscale{1.0}
\plotone{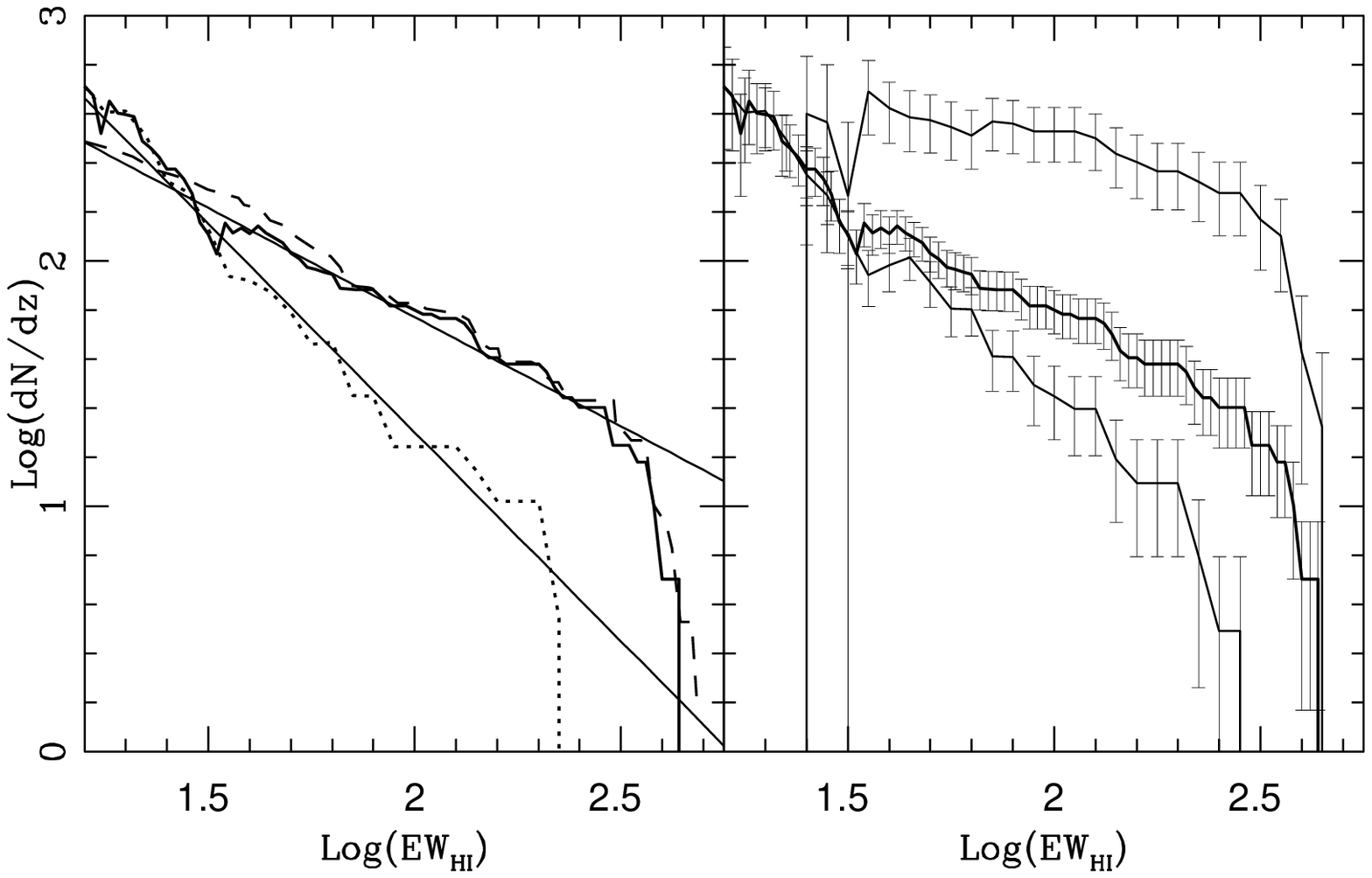}
\caption{Equivalent width distribution functions, expressed as the
logarithm of the cumulative line density as a function of the log of
the EW (in m\AA).  Panel $a$, an analysis of the derived mean EWDF
(heavy solid line), and that derived by PSSII (heavy dashed line), and
the EWDF for ${\cal C}({\cal T}^U)=0.01$ for $z < 0.036$ (dotted).
Thin lines are number-weighted linear fits to current data.  Note the
inflection in our mean EWDF (heavy line) at $\log(EW)=1.5$, which is
not seen in Penton \etal analysis, is continuous with the void EWDF.
Panel $b$, EWDFs with Poisson errors.  The mean distribution function
(heavy solid line), and the lighter lines are the EWDFs for catalogs
formed using ${\cal T}_{lim}=0.1$ as lower-bound (upper line) and
upper-bound (lower line).  }
\end{figure}

\subsection{Standard Processing}
Standard processing uses the ISOT model, accepts all 4-$\sigma$
clouds, and attributes galaxy mass based on apparent magnitude $m_B$,
a flat-$\Lambda$ model, and the TF relation pertinent to the $B$-band.
Fig. 9$a$ shows standard processing for all data irrespective of tidal
field (which we hereafter refer to as ``mean''), shown as the heavy
solid line.  Also shown, closely associated, is the mean EWDF taken
from PSSI (dashed line).  The two are very close except at ${\cal W}
\la 80$ m\AA; where the PSSI shows a broad bump, the present EWDF
shows an inflection to a higher slope.  This difference is due to the
different sensitivity functions employed in the present analysis.
Also shown (the dotted line) is the EWDF for ${\cal T} \le 0.1$, where
the data used excluded the range $z > 0.036$.  Notice that its slope
appears to be continuous with that of the inflected part of the mean
EWDF (solid line).  This is what one would expect if the low-EW part
of the mean EWDF is dominated by clouds in voids.  Also shown (thin
solid lines) are the number-weighted least squares linear fits to the
EWDFs.

In Fig. 9$b$ we show the Poisson errors associated with points for
the mean EWDF (heavy solid) and the EWDFs formed by the complementary
catalogs defined by ${\cal T}_{lim} = 0.01$; GDS is above, void is
below.  The error bars are not independent, but simply reflect the
number of clouds at that EW which are contributing to the EWDFs.  The
cutoffs evident in the EWDFs at high EW are probably due to the
tendency of large EW clouds to have associated metal-line absorption,
and so be rejected from the line lists.  The cutoff in the ${\cal
C}({\cal T}^L)$ EWDF (the upper line) at low EW is the result of the
paucity of low EW clouds in higher tidal field environments.  The
error bars show that the two complementary populations are
significantly distinguished.

The processing of the EWDFs was done for a series of ${\cal T}_{lim}$
equally spaced in $\log{\cal T}$.  Number-weighted (\ie, the number of
clouds contributing to the EWDF at a given EW) linear fits to the
equation,\be \log(d {\cal N}/dz)=\log{f({\cal W})}= C+S \log({\cal
W}/63), \ee are calculated, where $C$ is the intercept $\log(d{\cal
N}/dz)$ at ${\cal W}=10^{1.8} \simeq 63$ m\AA\ (the approximate median
$\log{\cal W}$) in the total cloud catalog, and $S$ is the slope.
Errors are derived from the residuals and the diagonal elements of the
variance-covariance matrix without number weighting.  These results
are presented in Table 2.  For comparison, the fits for the mean EWDF
are also presented.  Because of the break in the distribution apparent
at ${\cal W}\simeq 32$ m\AA, the fit is calculated for ${\cal W} \le
32$ m\AA, all ${\cal W}$, and ${\cal W} \ge 32$ m\AA, respectively.
Notice that the fitted parameters for the first resemble fits for void
EWDFs (${\cal T} \la 0.1$).
\begin{table}[h!] 
\begin{center}
\caption{Fitted slopes and intercepts as a function of ${\cal T}$ limits}
\medskip
\begin{tabular}{ccccc}
\tableline
\tableline

Mean EWDF & S & $\sigma_S$ & C & $\sigma_C$  \\
${\cal W} \le 32$ m\AA & -1.6670 & 0.0808 & 1.723 & 0.0345 \\
${\cal W} \ge 12$ m\AA & -0.9665 & 0.0285 & 2.082 & 0.0129  \\
${\cal W} \ge 32$ m\AA & -0.8900 & 0.0387 & 1.949 & 0.0166 \\
\hline
${\cal T}^U$ & S & $\sigma_S$ & C &  $\sigma_C$\\
0.010 & -1.4688 & 0.0583 & 1.7812 & 0.0227 \\
0.030 & -1.5277 & 0.0522 & 1.7204 & 0.0203 \\
0.100 & -1.5303 & 0.0557 & 1.7383 & 0.0217 \\
0.300 & -1.2823 & 0.0398 & 1.8233 & 0.0155 \\
1.000 & -1.1417 & 0.0535 & 1.8912 & 0.0235 \\
3.000 & -1.0267 & 0.0619 & 1.9450 & 0.0272 \\
\hline
${\cal T}^L$ & S & $\sigma_S$ & C &  $\sigma_C$\\
0.010 & -0.6690 & 0.0904 & 2.1087 & 0.0402 \\
0.030 & -0.5974 & 0.0939 & 2.2487 & 0.0417 \\
0.100 & -0.4845 & 0.1207 & 2.5602 & 0.0528 \\
0.300 & -0.5234 & 0.0970 & 2.7865 & 0.0429 \\
1.000 & -0.5515 & 0.1056 & 2.9507 & 0.0467 \\
3.000 & -0.3349 & 0.0969 & 3.0903 & 0.0433 \\
\hline
\hline

\tableline
\end{tabular}
\end{center}
\medskip
\normalsize
\end{table}

The data from the lower two subdivisions of Table 2 are plotted in
Fig. 10, where panel $a$ shows the trend of slope $S$ with $\log{\cal
T}$, and panel $b$ shows the intercepts $C$.  Slope fitting errors are
large compared to intercept errors.  The trends in Fig. 10 indicate
that in higher tidal field environments the redshift frequency of
absorption systems is larger, and the population more uniform in their
EWs, while for low tidal field environments the slopes are steeper
(more negative), representing an abundance of low EW clouds.  The
trend in the void slope with $\log{\cal T}$ shows signs of a relative
discontinuity somewhere between $0.1 \la {\cal T} \la 0.3$, the range
within which we hypothesize that the ``true'' void leaves off and the
GDS begins (\S5.6).  The same trend appears less obviously in the plot
of the intercepts (panel $b$).  If this trend is real it would suggest
that the deepest voids may have marginally more large EW clouds and
marginally larger line densities than in locations closer to void
edges.  Note also that choosing a larger lower tide limit (upper
sub-panels) is choosing smaller volumes around galaxies, and results
in significantly larger line densities (panel $b$).  The trend in the
line density in GDS is consistent with $C \propto {\cal T}^{0.43}$.
The tendency for large errors in attributed ${\cal T}$ to be
introduced in the GDS via positional errors makes this value highly
approximate.

\begin{figure}[h!] 
\centering
\epsscale{1.5}
\hspace{-1.0in}
\plotfiddle{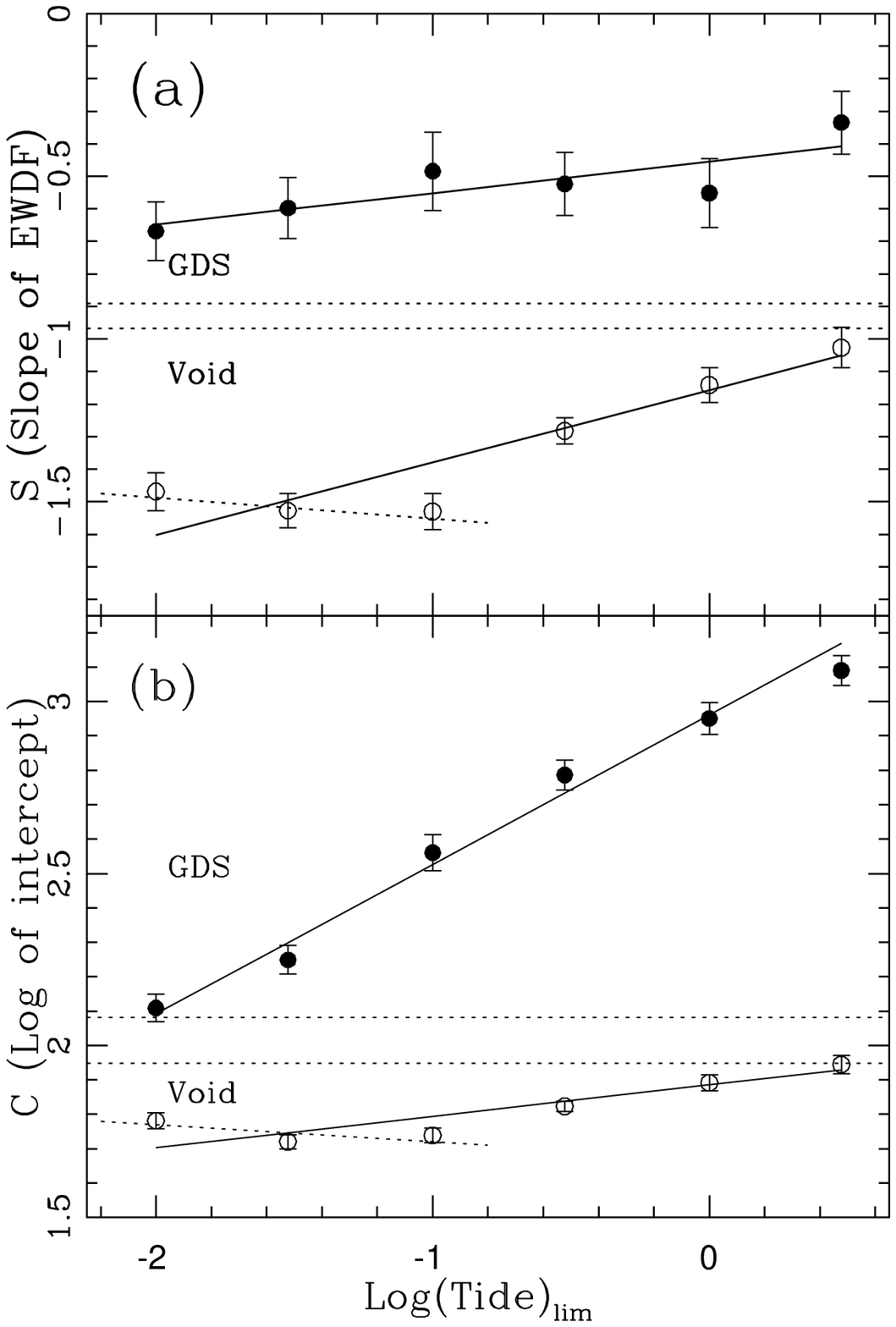}{5.25in}{0}{70}{70}{-302}{-97}
\caption{Standard model fits: The number-weighted linear fits to the
slopes $S$ (panel $a$) and intercepts $C$ (panel $b$), terms defined
in Eq. 45, are presented for various ${\cal T}$ limits of cloud
catalogs (see text).  The solid lines represent a linear fit to the
slopes (panel $a$) and intercepts (panel $b$) as a function of the
logarithm of the tide ${\cal T}$.  In panel $a$, the upper horizontal
dotted line has a value of the slope of the mean EWDF for ${\cal W}
\ga 32$ m\AA\ (the location of the inflection in the mean EWDF), and
the lower, ${\cal W} \ga 15$ m\AA.  Above that (filled circles) are
the EWDF slopes for catalogs ${\cal C}({\cal T}^L)$ (non-void), and
below the dotted line (open circles) are the slopes for ${\cal
C}({\cal T}^U)$ (void clouds).  In panel $b$, the horizontal dotted
lines represent the intercept (logarithm of $d{\cal N}/dz$) for the
mean EWDF for ${\cal W} \ge 32$ m\AA\ (lower) and for ${\cal W} \ge
15$ m\AA\ (upper).  Filled circles represent the intercepts of ${\cal
C}({\cal T}^L)$ EWDFs, and open circles represent the EWDF intercepts
for catalogs ${\cal C}({\cal T}^U)$.  The short dotted lines in the
void regions of panels $a$ and $b$ are linear fits to the points with
${\cal T}^U \la 0.16$ indicating a possible inflection of the trend of
slope and intercept with ${\cal T}^U$.}
\end{figure}

One may conclude that there is a significant variation in the EWDF
between the upper- or lower-limit tidal fields.  The systematic
variation of EWDF slope with ${\cal T}_{lim}$ is consistent with
expectations of the effects of tidal fields (\S3).  The modestly
slight variation in distribution functions for a range below ${\cal
T}^U \approx 0.16$ indicates that the void cloud EWDF is fairly
uniform over the $\approx 90\%$ of the universe occupied by voids.
The increasing concentration of clouds around galaxies, as shown by
the steady steep rise in the intercepts of catalogs ${\cal C}({\cal
T}^L)$ as ${\cal T}^L$ increases, appears consistent with infall of
clouds into the potential wells of galaxies and groups of galaxies.

\subsection{Error analysis by the variational approach} We need to
assure ourselves of the robustness of these results.  In previous
sections a number of possible biases, uncertainties and sources of
error were noted which could not be directly evaluated except by
noting the effect of different treatment of the data on the derived
EWDFs.  The results of these comparisons are presented below.

\subsubsection{ Completeness }
Recall (\S5.4) that the effect of incompleteness is to put a skew on
the depth of the luminosity function probed across a range of
redshifts, and hence to under-estimate ${\cal T}$ at the high-end of
the redshift range, relative to that at low redshift.  This effect is
shown graphically by the dotted lines in Fig. 4, where an absolute
magnitude upper-limit of $-19.8$ was imposed on the galaxy catalogs.  In
this section, the data is separated by a redshift constraint, and the
EWDFs are re-derived for the low-redshift part, and compared with our
standard.  We make our division at $ z=0.036$, at which point a
apparent magnitude limit of $m_B=16$ means that we are seeing all but
galaxies with ${\cal L} \le 0.74 {\cal L}^*$ ($v_c \simeq 145$ \kms).
Fig. 11 shows the results of that analysis in comparison to that of
the standard processing for catalogs based on ${\cal T}_{lim}=0.01$
and 0.1.  Figure 12 shows the linear fits for all ${\cal C}({\cal
T})$ for $z < 0.036$.  The lines are the fits to data at all redshifts
(Fig. 10), placed for the convenience of a visual comparison of the
changes.  Note that the slopes of void clouds have steepened; at
${\cal T}_{lim}=0.1$, the slope of EWDFs rises from $\sim -1.5$ to
$\sim -1.7$, possibly a better estimate of the true EWDF slope in
voids.  The previous impression of an inversion of the trend of slopes
for ${\cal T}^U \la 0.16$ (panels $a$ and $b$, short dotted lines from
Fig. 10) is now strengthened in significance (dashed lines).  Though
the number statistics are now weaker, it appears that we may have
gained in some respects by discarding higher-redshift data, for the
feature at ${\cal T} \simeq 0.16$ in Fig. 12 which was somewhat weak
in the standard processing using all the data now appears stronger
with the lower-redshift data.  This could be expected because fewer
clouds from higher tidal field environments may contaminate the low
tide catalogs when higher redshift data are culled.

\begin{figure}[h!] 
\centering
\epsscale{1.0}
\plotone{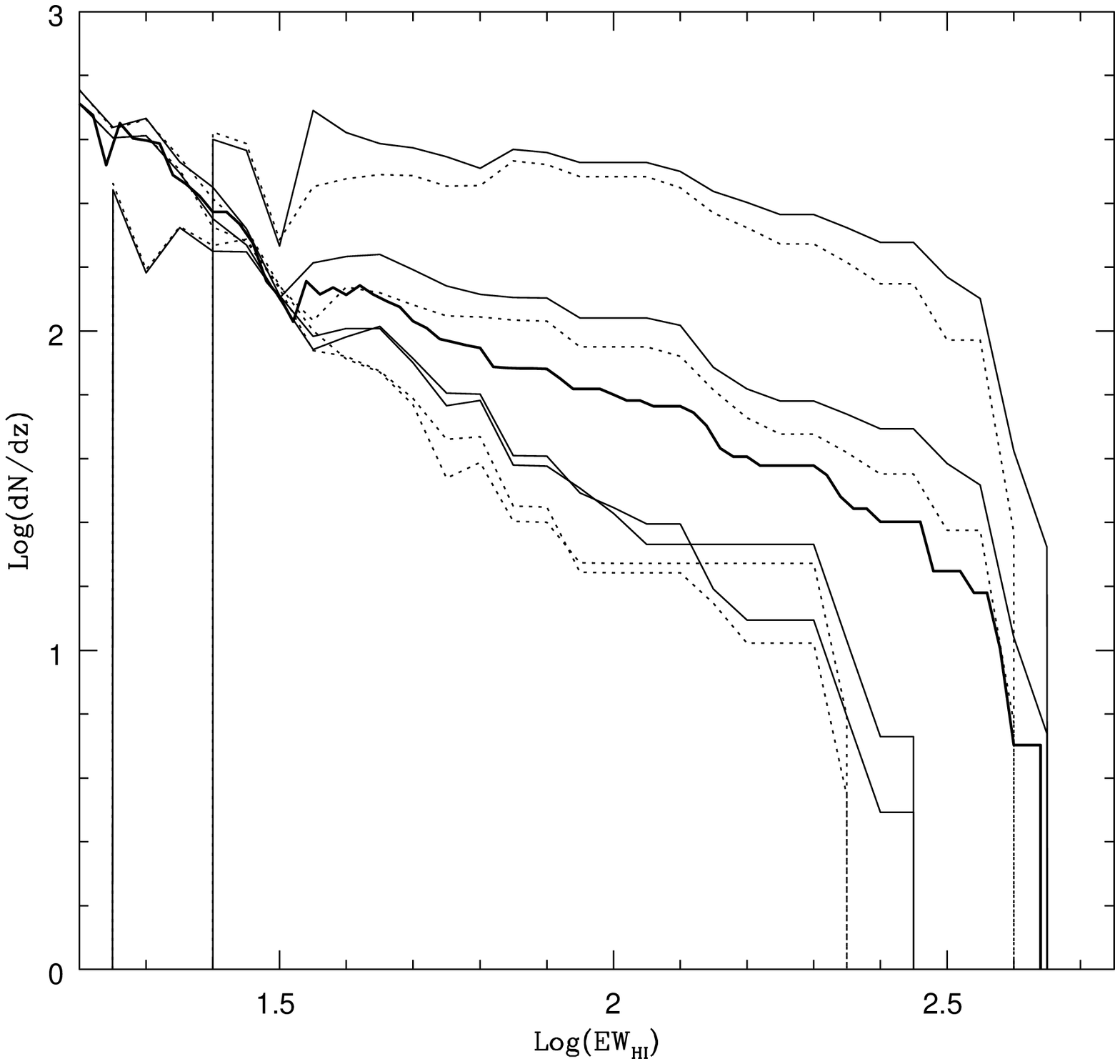}
\caption{Completeness test: The equivalent width distribution
functions (${\cal T}_{lim}=0.01$ and 0.1, with (solid) and without
(dotted), the redshift interval $z \ge 0.036$.  The median EWDF is
plotted (heavy solid line) for reference.  The ${\cal T}^U$ (void)
EWDFs are evidently somewhat steeper with the redshift cutoff; for
${\cal T}^U=0.1$, the slope steepened from --1.56 to --1.70.  The
${\cal T}^L$ EWDFs (upper pairs of lines) show a drop with the
rejection of higher redshift data.  This may be consistent with
evolution effects.}
\end{figure}

\begin{figure}[h!] 
\centering
\hspace{-1.0in}
\epsscale{1.5}
\plotfiddle{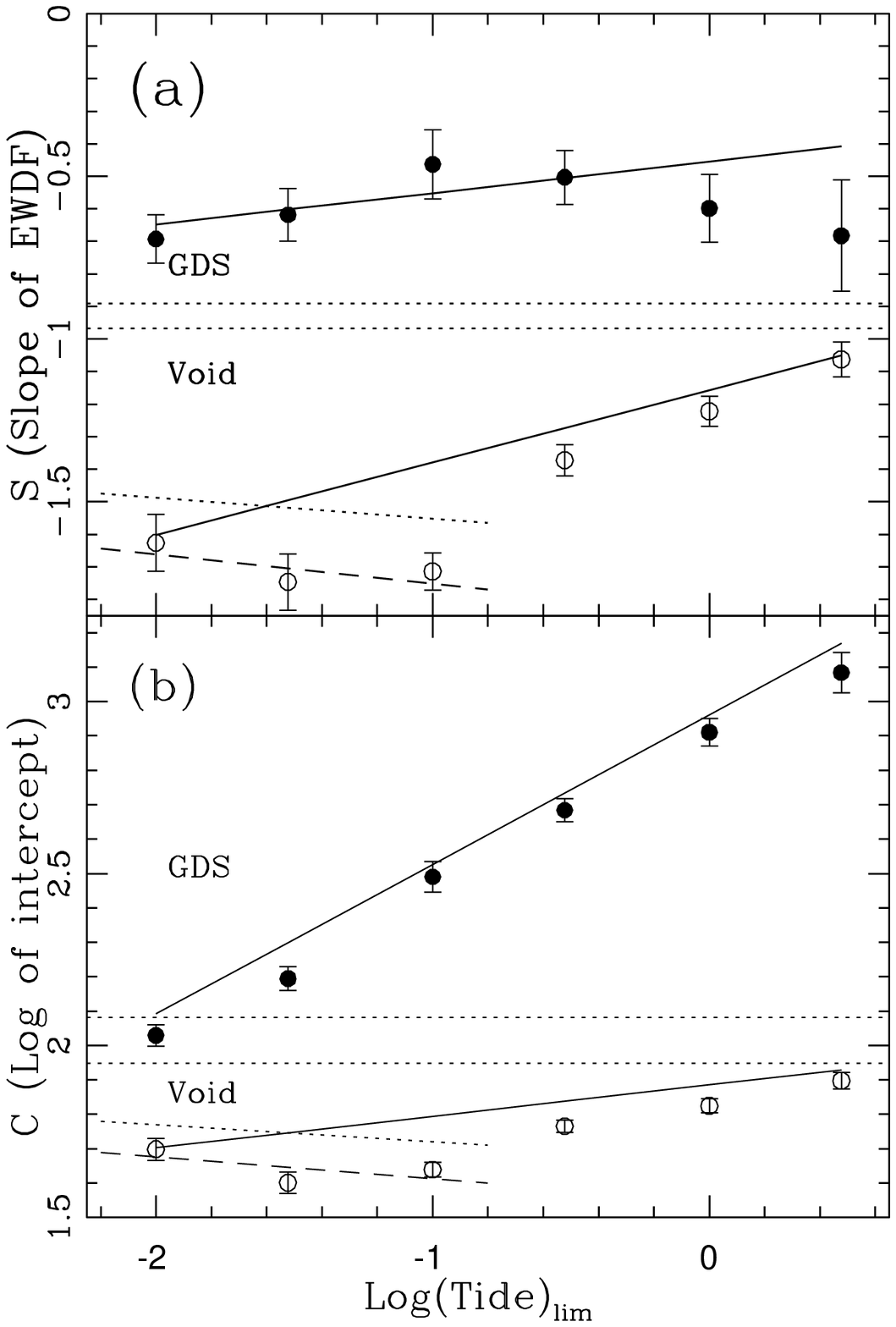}{5.25in}{0}{70}{70}{-302}{-97}
\caption{ Completeness test: Same as Fig. 10, except that slopes and
intercepts are fit to catalogs ${\cal C}$ constrained by $z \le
0.036$, and the solid lines are the previous fits found for catalogs
with no redshift constraint for comparison.  What appeared as a
marginal trend for increase in slope for GDS has now disappeared,
though the intercept relation remains relatively accurate (top
sub-panels of panels $a$ and $b$, respectively).  Note that what was
previously a marginal tendency for slopes and intercepts of void
clouds with ${\cal T}_{lim} \la 0.16$ (short, dotted lines) to have a
trend contrary to the general trend suggested by the linear fit, is
now more significant (short, dashed lines).  This provides evidence
for a relatively sharp change of character of cloud distributions near
${\cal T}_{lim} \simeq 0.16$ with the truncated isothermal model. }
\end{figure}

\subsubsection{ Effect of early-type galaxies }
Recall (\S5.1) that in attributing mass to galaxies the TF relation
was used to transform the absolute magnitudes into galaxy masses.
Early-type galaxies do not follow the TF relation, however, they do
follow a quite similar relation, basically ${\cal L} \propto
\sigma^{\sim 4}$.  This is steeper than the TF relation, but the main
effect, for our purposes, is that the mass-luminosity ratio is larger
for early-type galaxies.  This can be checked by alternatively using,
or not using, an altered attributed mass for those galaxies listed in
the CfA catalog as having negative de Vaucouleurs T-Types (those are
generally lenticular or elliptical galaxies).  The alteration involves
multiplying the normally attributed mass by a factor of 3.0 when the
T-Type is less than or equal to zero.  Since ellipticals are more
often found in clusters, we might expect that this would attribute
more masses to large groups, or clusters of galaxies, inflating the
${\cal T}^L$ contours somewhat, which would reduce the EWDF for the
GDS at a given ${\cal T}^L$.  There should be relatively little effect
on the void EWDFs since clouds are thought to avoid high density
regions \citep{Morris:91,Morris:93} where early-type galaxies are
concentrated.

\begin{figure}[h!] 
\centering \epsscale{0.9}
\plotone{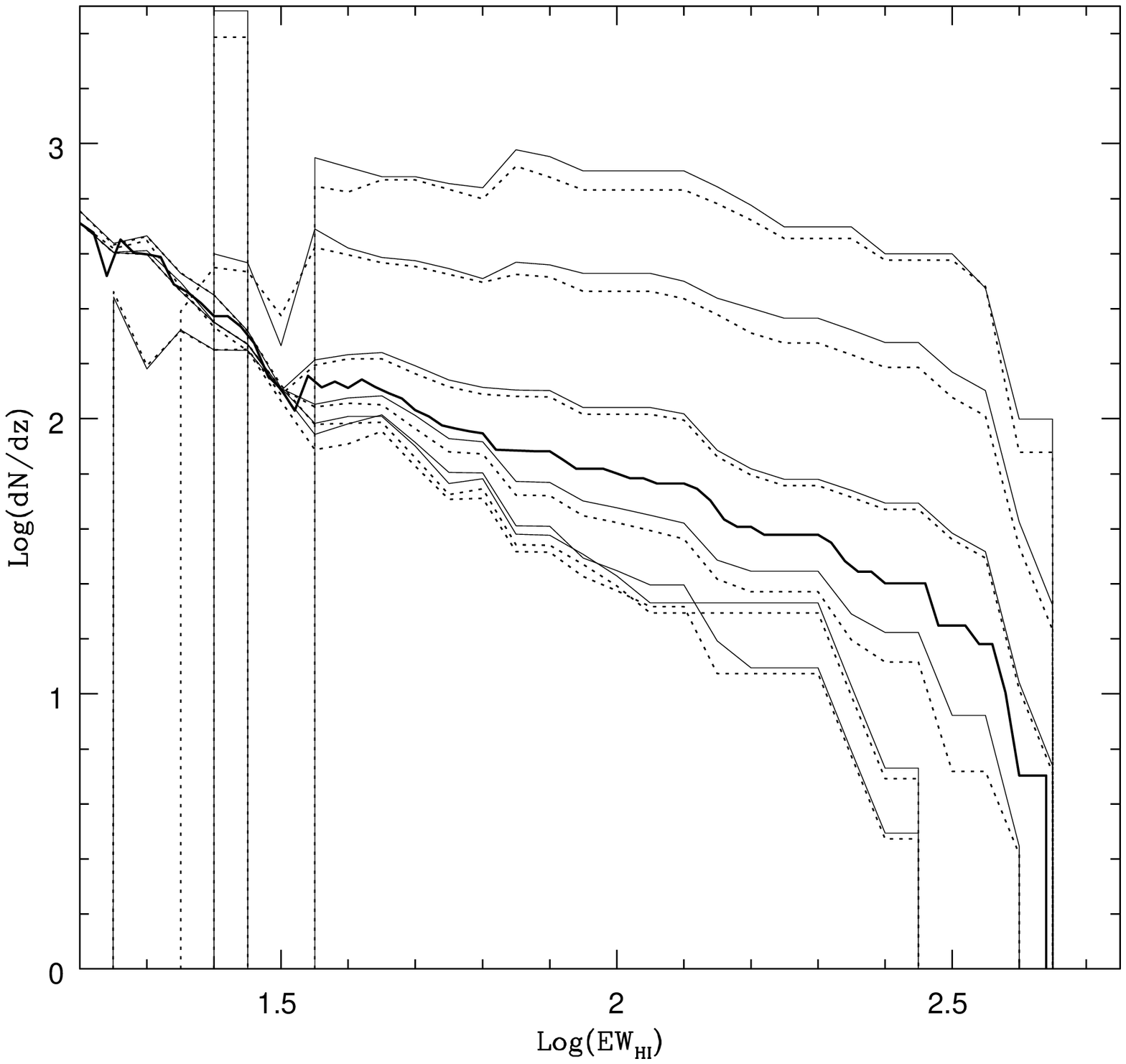}
\caption{The T-Type test: The equivalent width distribution functions
with (dotted lines), and without (solid), the T-Type corrections for
early-type galaxies, which makes the attributed galaxy mass 3 times
greater.  Tide limits are ${\cal T}_{lim}=0.01$, 0.1, 1.0.  The
tendency is for EWDFs based on tidal field lower limits to be more
influenced in high tidal field, than low tidal field environments,
though only marginally so.}
\end{figure}

As can be seen in Fig. 13, there is slight reduction in the line
density of the DFs.  The reason for this is probably that the added
mass simply increases the volume of space at a given tidal field
limit, which then has a slightly lower EWDF, as noted in Fig. 10$b$.

\subsubsection{Effects of halo type}
The NFW halo is significantly less massive than the truncated
isothermal halo (\S5.1, and Fig. 3), and therefore when halos of
galaxies are interpreted as NFW, the main effect is that the tidal
fields will be lower by a factor of $\sim 5$.  Thus we can expect that
the intercepts and slopes will be shifted to lower $\log{\cal T}$ by
a factor of $\approx 0.7$.  These results are presented in the form of
fitted slopes and intercepts of number-weighted linear fits in
Fig. 14.
 
\begin{figure}[h!] 
\centering
\epsscale{1.5}
\hspace{-1.0in}
\plotfiddle{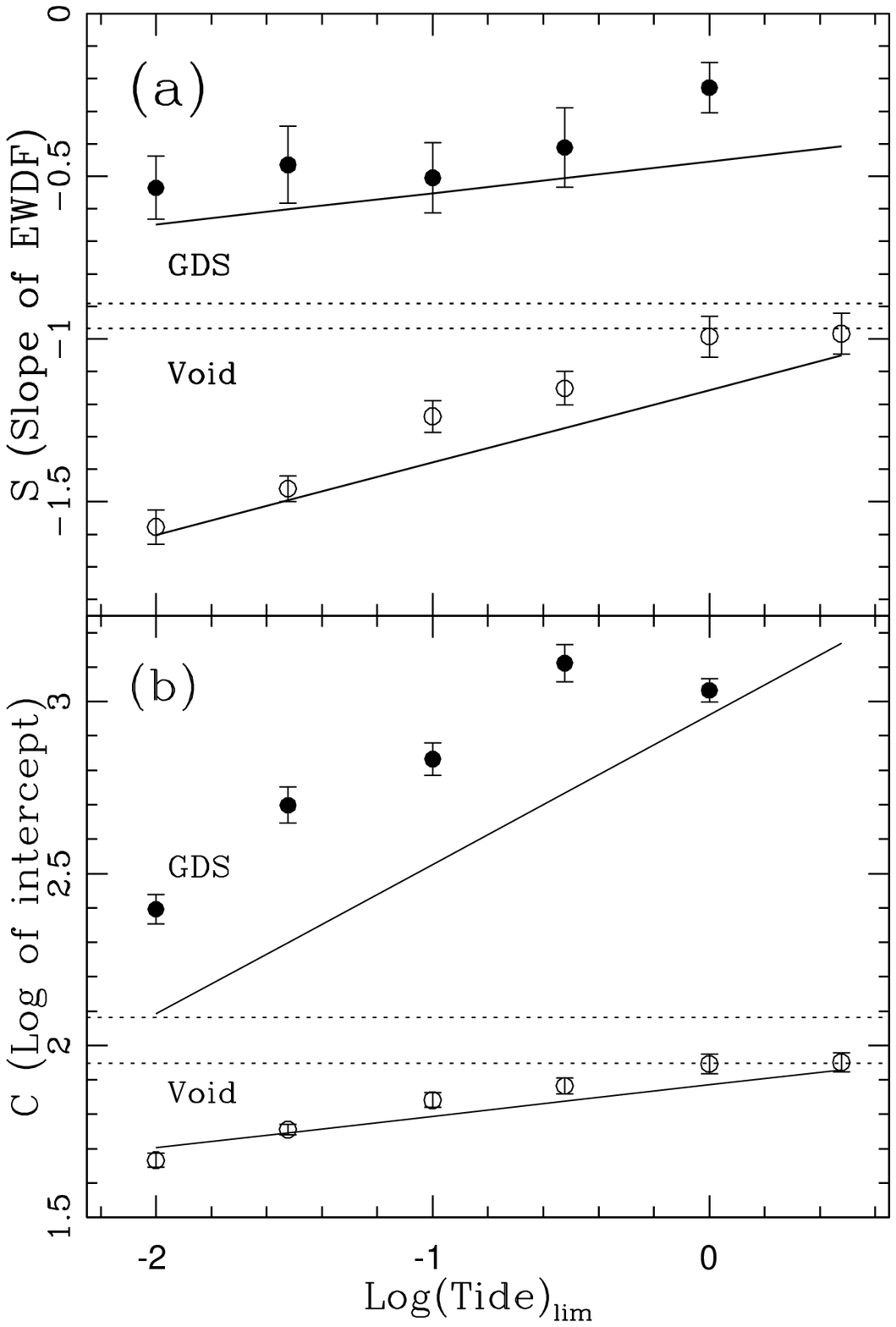}{5.25in}{0}{70}{70}{-302}{-97}
\caption{ Halo-type test: Same as Fig. 10, except that slopes and
intercepts are fit to catalogs ${\cal C}$ with tides generated by
halos interpreted as NFW.  The solid lines are the previous fits found
for catalogs with isothermal halos.  To first order the points are
displaced by $\sim 0.7$ to the left (in $\log{\cal T}$) relative to
the points in Fig. 10 for the isothermal halo due to the smaller NFW
halo mass. }
\end{figure}

It is difficult to judge from Figs. 10 and 14 alone which halo is most
likely to be more correct.  However, we saw signs in Fig. 8. on the
basis of most probable b-values, that the transition from void to GDS
clouds occurs in the range $0.1 \la {\cal T} \la 0.3$.  A $\log({\cal
T})$ offset of -0.7 would place this transition at ${\cal T} \approx
0.03$ ($\log{\cal T}\simeq -1.5$).  While neither of these tides could
not reasonably be responsible for truncating the clouds in the
transition zone, at $z=1$, a lookback time of 57\% of the age of the
universe, the expected tidal field is $\sim 8$ times greater, which
according to Eq. 19, constrains stable clouds to have densities
greater than $\sim 2.5$, and $0.48 \rho_{crit}$ for isothermal and NFW
halos, respectively.  However, it does not seem reasonable to expect
that the GDS and void clouds could be distinguished at a level as low
as 0.5 $\rho_{crit}$ level; clouds of such low density could not
reasonably be expected to be detected with the current technology.
Such a cloud in the Hubble flow would produce a column of only
$10^{12} ~\cmtw$ (${\cal W} \la 4$ m\AA) over a velocity range of
$\sim 220 ~\kms$.  Hence, it appears that gas at a density of $\sim
2-5 \Omega_b \rho_{crit}$ is being stripped at the void-GDS boundary,
where the tidal fields by themselves, even in the last few Gyr, could
not reasonably be expected to be effective, unless galaxy halos are
significantly more massive than even the isothermal model used here.

\section{Discussion}

Numerous 3-D cosmological simulations have suggested that voids have a
low density, and would present very few absorbers above present-day
obsesrving limits.  However, careful separation of clouds into
catalogs according to the ambient tidal field has resulted in the
discovery of a void cloud population that is by no means
insignificant.  The line density of void clouds reaches a value
$d{\cal N}/dz \approx 500$ for ${\cal W} \ge 10^{1.2}\approx 15.8$
m\AA.  An order of magnitude calculation may convey the meaning of
this.  Equation 4 shows that at $z=0$, and $r_p=50$ kpc, the comoving
density $n \simeq 16 ~{\rm Mpc}^{-3}$, and the mean free path between
clouds ${\widehat l}\simeq 8$ Mpc, or 2.4 \AA.  An average cloud mass
of $10^9 ~\Msun$ (a reasonable mass for a just-detectable truncated
isothermal halo of circular velocity $v_c\simeq 11 ~\kms$) implies
$\Omega_m=0.1$ in voids.  However, an inter-cloud medium of unknown
density must exist which may add substantially to the density in
voids.

Supporting the contention that these are actually void clouds is the
clear distinction between the slopes of EWDFs based on the
complementary catalogs ${\cal C}$ distinguished by limiting tidal
fields ${\cal T}_{lim}$, and the difference in the most probable
$b$-values of their clouds.  Under the truncated isothermal halo
model, the transition from void to GDS environments occurs at
$log{\cal T}\approx0.8 \pm 0.2$, or ${\cal T} \approx 0.16$.  Not only
do the slopes and intercepts of the void EWDFs undergo a strong change
(Figs. 10, 12), but the differential histogram of cloud $b$-values
also shows the transition.  At this tidal field limit, $\sim90\%$ of
the universe is void, and 10\% is GDS.

That the change in cloud character noted above can be associated with
a certain tidal field limit under a given assumed halo type supports
the contention that tidal fields are a relevant quantity, associating
cloud stability and cloud characteristics with void and GDS
environments.  It further supports the contention that clouds are
self-gravitating to some extent, for the variation of EWDF with tidal
field is consistent with the predicted effects on such clouds.  In the
case of void clouds, the lack of strong tides allows a distribution of
clouds which is unconstrained to the current detection limits,
allowing small, delicate structures to exist.  However, their true
nature cannot be known without successful modeling of them in
cosmological simulations.  In the case of GDS clouds, the tidal
truncation which galaxies and groups of galaxies impose on clouds
appears to have resulted in a paucity of low-EW clouds, and a general
flattening of the EWDFs.  On the other hand, if GDS environment is
well-characterized by the filamentary structures seen in 3-D hydro
simulations, and have a relatively high density and temperature, it is
possible that much of the observed truncation is by ram pressure and
shear forces as clouds fall into the GDS from voids.  This would solve
the problem (noted in \S6.2.3) of tidal field levels at the location
of the apparent transition between void and GDS being too low with
either halo model to easily reconcile with the probable density of gas
being stripped from clouds moving from void to the GDS.  Both effects
have a common source in the gravitational potential of the mass
concentration, and a similar result in the stripping of low-density
gas, so that even if ram-pressure stripping is a major factor in the
difference between void and GDS clouds, the tidal field is still an
effective and valid way to separate void and GDS clouds.

There is a hint of structure in voids -- deep voids appear to have
marginally larger line densities, and shallower EWDF slopes; a hint of
more massive clouds there.  This may suggest some degree of
hierarchical clustering has taken place.

In comparing ISOT and NFW models, the most obvious difference is that
in mass (NFW is a factor of 4 to 5 smaller).  The NFW halo has a
weaker ability to tidally truncate clouds, and to maintain the filling
factor of 0.1 for GDS, one requires that ${\cal T}_{lim}\simeq 0.03$
has in the past tidally truncated the GDS clouds (see \S6.2.4).  But
this may be too small to reasonably have tidally truncated those
clouds.  In addition, the scaling of the absorption radius with
luminosity in the $B$-, and $K$-bands \citep{Chen:01} agree nicely
with the scaling relations of ISOT (Eq. 29), but do not fit well with
the NFW halo (Eq. 40 -- see \S\S5.1.1, 5.1.2).

Though I consider these general findings to be secure, many
improvements could be made in the quality and quantity of data.  For
instance, the depth and quality of the data in the CfA redshift
catalog leaves much to be desired.  However, when the galaxy catalogs
from the Sloan Digital Sky Survey become available, much of these
problems will be solved; both deeper ($m_{g^*}=17.65$ mag), more
consistent photometry, and generally superior redshifts will be
available, so that tidal fields may be calculated to greater accuracy.

The quantity of absorption systems, and the sensitivity of the
observations also leaves something to be desired.  However, there will
soon be published results of Space Telescope Imaging Spectrograph (STIS)
observations of $\sim120$ new absorbers (M. Shull, priv. comm.), more than
doubling the current number of absorbers.  Somewhat farther in the
future, the Cosmic Origins Spectrograph (COS) will provide
significantly deeper spectrography of low redshift AGN, and should
greatly increase the number of \hone\ absorbers available for study.

While this analysis has displayed the nature of the distinction
between void, and GDS clouds, we know little about the clouds
themselves.  The EWDF is just a sampling of the clouds weighted by
cloud cross-section at a given EW.  We do not know their mass, their
cross-section, or their density.  We have a tentative cloud model --
clouds of primordial gas whose expansion is restrained by dark matter
halos -- but do not know in any detail what parent population has
given rise to these absorbers, nor what contribution they make to
$\Omega_m$.  Though analytical models may help to clarify what the
current state of clouds might be, the accurate implementation of the
cloud hypothesis adopted here requires a hydrodynamical study
beginning at the point of reionization in order to assess the current
extent and dynamical characteristics of void clouds under a variety of
halo models.  These results could be used to substantiate, or
invalidate, those models.  If the former, it would help determine the
number and mass-distribution of clouds which are consistent with the
observed void EWDF, and thus estimate their contribution to
$\Omega_m$.  This task remains for a paper now in preparation.

\section{Conclusions}
The large line density of clouds in voids, where simulations have
predicted very few clouds at EWs accessible to even STIS observations
(R. Dav\'e, priv. comm. 2001), appears to require massive dark halos
to restrain the dispersal of gas needed to produce detectable
absorbers.  We have seen that such halos are consistent with the
general predictions of hierarchical structure formation if small halos
did not generally form Population III stars.  We further find that the
increase in line density of clouds with increasing tidal fields in GDS are
consistent with continuing infall of clouds from void regions into the
GDS.

We make the following specific conclusions.
\begin{enumerate}

  \item Tidal fields are a convenient and effective parameter with
  which to separate clouds into the complementary environments;
  ``void'', and ``galaxy dominated'' spaces.  However, it is possible
  that the tidal field stripping of clouds is significantly
  supplemented in GDS by ram-pressure stripping by the diffuse,
  ambient dissipated gas populating filamentary structures which
  appear in 3-D cosmological simulations.

  \item The EWDF is significantly steeper in voids than the median
  EWDF, suggesting that the void cloud population has most of its mass
  at smaller scales.

  \item Most probable void cloud Doppler parameters are
  characteristically half that of GDS clouds.

  \item The line density in GDS is of order more than 10 times larger
  for ${\cal W} \ge 63$ m\AA, but for ${\cal W} \la 35$ m\AA, void
  clouds dominate the mean EWDF.

  \item The void cloud EWDF is essentially constant over $\sim 90\%$ of the
  universe, while the GDS EWDFs have line densities that increase
  dramatically with increases in ${\cal T}^L$.

\end{enumerate}


We have analyzed the void-cloud population and found significant
differences between that and the picture provided by 3-D hydrodynamic
simulations.  It is speculated that the lack of resolving power in the
simulations may be partially responsible for this difference.  However,
it appears that even so, the sheer number density of clouds which must
be in voids is startling, and may presage a shift in thought about
\lya\ clouds.

I am very grateful to Hy Spinrad, who provided support for this
research, and helpful suggestions.  I am also grateful for the many
comments and helpful suggestions from Christopher F. McKee.  I thank
Steve Penton and J. Michael Shull for their help with the sensitivity
functions and cloud identification issues.  Finally, I wish to thank
the anonymous referee for many helpful suggestions and comments.

\begin{center}
APPENDIX

TRUSTWORTHINESS OF THE CfA REDSHIFT CATALOG
\end{center}
\setcounter{equation}{0}
A lot will depend on the reliability of the redshifts and magnitudes
in the assembled catalogs.  Since the CfA galaxy catalog is a
compilation of data from many sources, its tabulated magnitudes are
somewhat inhomogeneous.  To check the validity of these magnitudes,
the galaxy positions are used to query the Automated Plate Scanner
Facility at the University of Minnesota (\eg,
\citet{Pennington:93})\footnote{ The APS databases are supported by
the National Aeronautics and Space Administration and the University
of Minnesota. The APS databases can be accessed at
http://aps.umn.edu/.}.  The APS is based on digitized Palomar
Observatory Sky Survey (DPOSS) plates.  Not all galaxies were
``found'' by the APS.  Reasons were often that the relevant plate was
not mounted.  Sometimes the galaxy was simply not found, though STScI
digitized sky survey images\footnote{The Digitized Sky Surveys were
produced at the Space Telescope Science Institute under the
U. S. Government grant NAG W-2166, and may be accessed at
$http://archive.stsci.edu/cgi-bin/dss_form$} clearly showed their
presence (often these were edge-on spirals).  In a very few cases,
nothing was found.  Roughly half of the CfA galaxies were matched by
an APS galaxy.  The apparent magnitudes are converted to absolute
magnitudes, corrected for galactic extinction \citep{Schlegel:98}, and
inclination \citep{Tully:98}.  A scatter plot makes it clear that some
of the magnitudes were anomalous.  However, the overwhelming majority
of galaxies fit the relation, $M_{APS}=M_{CfA} + (0.65 \pm 0.5)$ mag,
with a slope of 1.004

\begin{figure}[h!] 
\centering
\setcounter{figure}{0}
\epsscale{1.0} 
\plotone{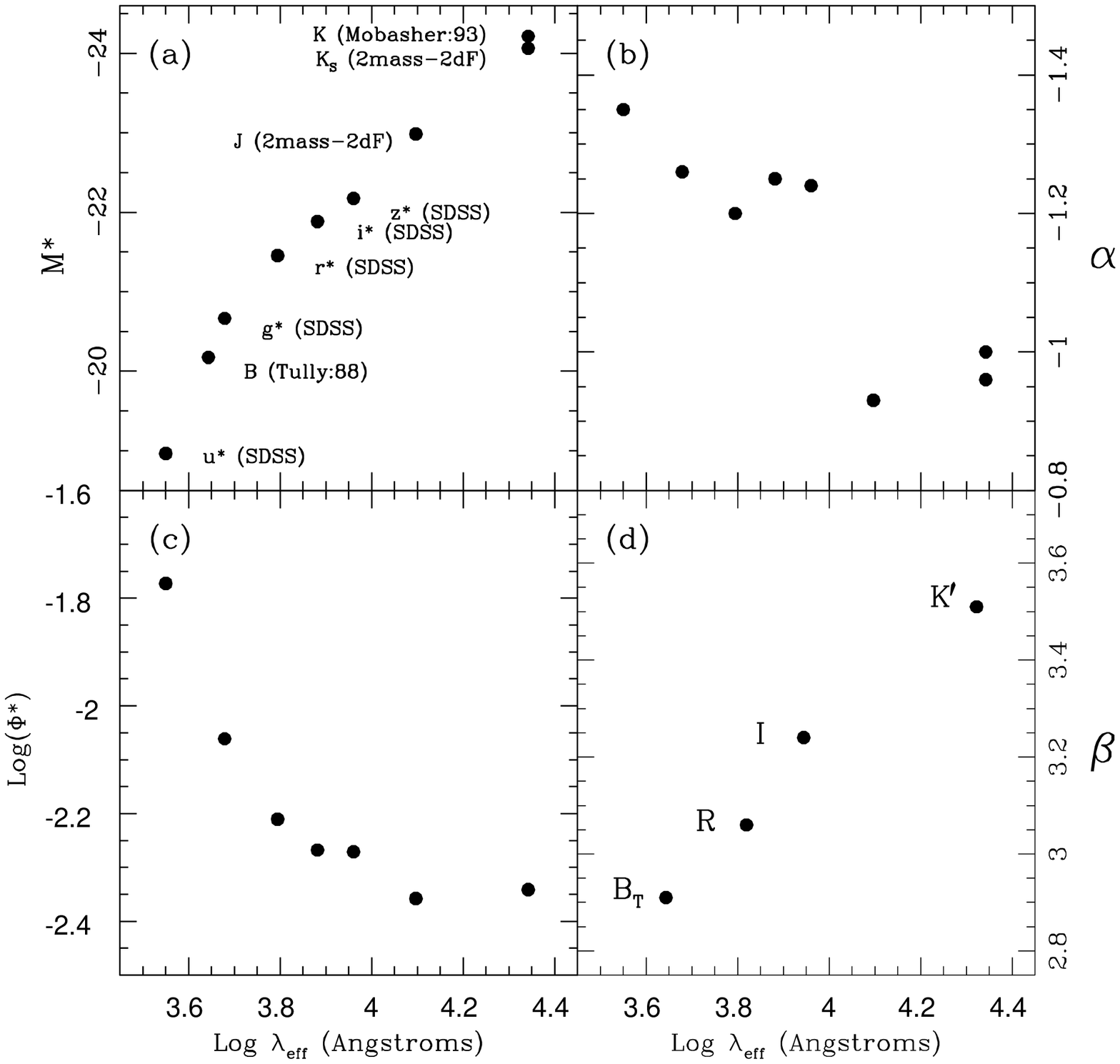}
\caption{Panels (a) through (c) show luminosity function parameters.
Panel (a) shows $M^*$ as a function of various spectral bands using
$h=0.75$ (\citealt{Blanton:01,Cole:01,Mobasher:93,Tully:88}).  Panel
(b) shows the faint-end slope $\alpha$ as a function of
$\lambda_{eff}$.  Panel (c) shows the log of the normalization
$\Phi^*$ with $h=0.75$.  Panel (d) shows the slope $\beta$ (see
Eq. 26, \S5.1) of the Tully-Fisher relation as a function of spectral
band \citep{Tully:00}.}
\end{figure}

The stated error of the APS magnitudes for integrated isophotal galaxy
magnitudes \citep{Pennington:93} was estimated to be $\sigma_{APS}
\simeq 0.3$ to $0.5$ mag, but a more recent estimate is 0.3 mag
(R. Humphreys, priv. comm., 2001), reflecting the added
sophistication of their processing.  The observed dispersion is 0.5
mag in the APS-CfA plot, which is also the dispersions of APS and CfA
magnitudes combined in quadrature.  Thus, we estimate $\sigma_{CfA}
\simeq 0.4$ mag.

The 0.65 mag offset indicates that the APS magnitudes are fainter.  We
shall see that much of this offset can be understood in terms of the
different spectral band in which the magnitudes are measured.  The CfA
magnitudes are generally B-band magnitudes with $\lambda_{eff} \simeq
4200-4400$ \AA.  The APS magnitudes are POSS O-band with
$\lambda_{eff}\simeq 4000$ \AA.  A compilation of published studies of
luminosity functions (LF) in a range of spectral bands
(\citealt{Blanton:01}; \citealt{Cole:01}; \citealt*{Mobasher:93};
\citealt{Tully:88}; \citealt{Tully:00}) show how the defining
parameters of the fit vary with the spectral band, revealing a tight
relation between $\phi^*$, and $M^*$ (Fig. A1a, c) as a function of
spectral band.  Let us analyze the CfA \vs the APS magnitudes in the
light of this figure.  Note first that as one chooses bluer bands, the
galaxy luminosity is smaller at the ``knee'' of the luminosity
function, and the normalization $\phi^*$ is larger.

It is the Tully-Fisher relation \citep{Tully:98} which allows us to
use the LF data to get our answer.  TF can be
expressed,
$$ M_{band}=M_{band}^* - 2.5\left[\beta_{band}\left\{\log(2v_c)-\log(2
v_0) \right\}\right].$$ Applying this once each to $O$ and $B$
magnitudes and subtracting them, we have the transform from $O$ to $B$
for a galaxy of circular velocity $v_c$, \be
B=O-(O^*-B^*)+2.5\left\{\beta_0-\beta_B \right\}
\log\left(\frac{v_c}{v_0}\right), \ee where $\beta_{band}$ is the
Tully-Fisher slope (see Fig. A1d), and $v_0$ is a fiducial velocity of
158 \kms.  A linear fit of the $B_T$, $R$, and $I$ bands (Fig. A1d)
implies $\beta_0 \simeq 2.83$, and $\beta_B\approx 2.87$ to 2.89.
$M^*$ and $\phi^*$ change relatively fast in these bands
($M_O^*=-19.0$ mag, $\phi_O^* \simeq 0.029 \h^3 \,{\rm Mpc}^{-3}$ and
$M_B^*=19.3-19.7$ mag, $\phi_B^* \simeq 0.026-0.023 \h^3 \,{\rm
Mpc}^{-3}$) for the 4200 \AA\ and 4400 \AA\ versions of $B$,
respectively.

Applying these values of $M^*$ to Eq. 1 shows a shift of $+0.4$ to
0.7 magnitudes for a galaxy with $v_c=158 ~\kms$ going from $B$ to $O$
band magnitudes; $O$ is ``fainter''.  For galaxies with $v_c$ smaller
than 158 \kms, there is only a slight skew in the sense that for small
galaxies, the shift in attributed absolute magnitude is slightly
larger.

One can conclude that the CfA magnitudes are generally consistent, and
trustworthy within the constraints established above.

\bibliographystyle{apj} 

\end{document}